\documentclass{article}
\usepackage[english]{babel}
\usepackage[resetlabels,labeled]{multibib}
\usepackage{amsmath}
\usepackage{adjustbox}
\usepackage{caption}
\usepackage{subcaption}
\usepackage{float}
\usepackage{lscape}
\usepackage{rotating}
\usepackage[letterpaper,top=2cm,bottom=2cm,left=3cm,right=3cm,marginparwidth=1.75cm]{geometry}
\usepackage{tablefootnote}
\usepackage{amsmath}
\usepackage{graphicx}
\usepackage[colorlinks=true, allcolors=blue]{hyperref}

\title{Exploring the Dynamics of the Specialty Insurance Market Using a Novel Discrete Event Simulation Framework: a Lloyd's of London Case Study}

\author{Sedar Olmez\thanks{Joint First Authors}, Akhil Ahmed\footnotemark[1], Keith Kam, Zhe Feng, Alan Tua\thanks{Corresponding Author: \texttt{alan.tua@ki-insurance.com}}}

\begin{document}
\maketitle

\begin{abstract}
This research presents a novel Discrete Event Simulation (DES) of the Lloyd's of London specialty insurance market, exploring complex market dynamics that have not been previously studied quantitatively. The proof-of-concept model allows for the simulation of various scenarios that capture important market phenomena such as the underwriting cycle, the impact of risk syndication, and the importance of appropriate exposure management. Despite minimal calibration, our model has shown that it is a valuable tool for understanding and analysing the Lloyd's of London specialty insurance market, particularly in terms of identifying areas for further investigation for regulators and participants of the market alike. The results generate the expected behaviours that, syndicates (insurers) are less likely to go insolvent if they adopt sophisticated exposure management practices, catastrophe events lead to more defined patterns of cyclicality and cause syndicates to substantially increase their premiums offered. Lastly, syndication enhances the accuracy of actuarial price estimates and narrows the divergence among syndicates.  Overall, this research offers a new perspective on the Lloyd's of London market and demonstrates the potential of individual-based modelling (IBM) for understanding complex financial systems.
\end{abstract}

\section*{Acknowledgements}
A special thanks to the Alan Turing Institute's Internship Network (TIN) in supporting the project. Accenture for their partnership with academia. Ki and Brit syndicates for bridging the relationship between academia and industry, providing the resources to undertake the research project. A special thanks to Reuben Thomas-Davis for developing and providing access to the HADES framework.

\section{Introduction}
The specialty insurance market is a large-scale complex system with many uncertainties, complex business relationships and non linear dynamics and interactions amongst participants. To quantify the behaviour of this complex system, researchers have turned to various traditional approaches ranging from time-series methods \cite{Owadally2019AnMarkets, Owadally2018TheCrises, Piotr2016UnderwritingCrises, Venezian1985RatemakingInsurance, Weiss2004U.S.Capacity, boyer2012} to differential equation based mathematical models \cite{Einav2010BeyondMarkets, Cummins1992FinancialInsurance, Norberg1995AInsurance}. This top-down, aggregate approach of modelling complex systems may fail to capture the interactions which occur at a micro-scale which ultimately lead to large-scale emergent phenomena. Indeed, many researchers attest to this challenge within the specialty insurance modelling literature. For instance, the ``Underwriting Cycle", an important phenomena where the market undergoes periods of high profitability, less competition and low profitability, high competition among insurers (in the insurance literature known as the hardening and softening of the market), is seldom accurately captured using the vast number of traditional aforementioned methods. Boyer et al. demonstrates the challenges in modelling this phenomena in their paper \cite{boyer2012} where they show time-series methods fail to model this important stylised fact of the specialty insurance market. All of the above points reflect the need for an alternative approach, this is where Individual-Based Modelling (IBM) approaches can help.

Individual-Based Models, which encompass modelling frameworks such as Discrete Event Simulations (DES) to Agent-Based Models (ABMs), can alleviate the above challenges posed by traditional methods utilised in the study of insurance markets. For example, these models allow complex systems to be built from the bottom-up, focusing on interactions at the micro-scale between autonomous ``agents" which can represent anything from people, organisations to more granular entities such as molecules. Since the late 90's, IBM applications have grown in popularity, starting from computational social science to ecological studies, biology and environmental sciences. Given its influence in various research fields, the IBM method continues to advance across disciplines, to more relevant areas of research including the modelling of economic markets such as housing, insurance and energy \cite{Owadally2018TheCrises,Arthur2006Out-of-equilibriumModeling, Hamill2015Agent-BasedEconomics, Owadally2019AnMarkets, Farmer2009TheModelling, Heinrich2021AHomogeneity, Johnson2014StrengtheningModeling, 
Groff2019StateOverview, Baptista2016MacroprudentialMarket, Ge2017EndogenousMarket, Yun2020HousingDebt-to-income, Heppenstall2007GeneticMarket, McLane2011TheManagement}. 

The IBM methodology, can enable researchers to observe the emergent phenomena at the aggregate level which arise from the micro level interactions of autonomous heterogeneous agents. Each type of agent can embed behaviours and actions reflective of real-world concepts. In fact, these powerful IBM methodologies have reached industry practitioners within the wider retail insurance market, evidencing the appeal of such approaches in modelling market dynamics. The Institute and Faculty of Actuaries (IFA), the professional body and regulator of actuaries in the UK, have gone so far as to start an ``Agent-Based Modelling Working Party" where workshops have taken place, and research such as \cite{Zhou2013ApplicationCycles} presented. Moreover, the Bank of England has published working papers utilising the approach \cite{Baptista2016MacroprudentialMarket}. On the other hand, academics from the Bayes Business School \cite{Owadally2018TheCrises, Owadally2019AnMarkets} and The Institute for New Economic Thinking \cite{Farmer2009TheModelling, Farmer2022Agent-BasedFuture, Heinrich2021AHomogeneity} have all applied ABMs to the insurance market with significant success.

Given the spotlight on the modelling of insurance markets and a lack thereof in specialty insurance markets, this paper will utilise the findings from the existing ABM insurance literature. We combine these findings with the expert knowledge gained through workshops and interviews with underwriters, actuaries, capital modellers, portfolio analysts, brokers and algorithm engineers at Ki and Brit syndicates (insurers) within Lloyd's of London (the oldest specialty insurance marketplace). These findings allow us to create a novel, discrete event simulation (DES) to study the emergent properties and drivers of the specialty insurance market at various spatio-temporal resolutions. The model proposed in this research article is novel in several aspects: 
\begin{itemize}
    \item The model utilises a DES framework open-sourced by Ki insurance (HADES). HADES consists of two main components:
    \begin{itemize}
        \item Processes - these components are responsible for performing some actions and subsequently emitting events as a response to input events, i.e., Broker process responding to Day, Month or Year events.
        \item Events - a piece of information that can be altered given its interactions with processes.  
    \end{itemize}
    \item The model incorporates functionality for different event types, such as catastrophe losses (typically low-frequency, high-severity events) and attritional losses (high-frequency, low-severity events). While previous research endeavours have only sought a single type of loss.
    \item Interviews and workshops conducted with experts within Ki and Brit insurance have shaped the conceptualisation of the agents represented in the model.
    \item The model simulates the dynamics of the Lloyd's of London specialty insurance market by incorporating unique features such as lead and follow insurers.
\end{itemize}

This research article will conduct four experiments with varying complexity to demonstrate the emergent properties of the specialty insurance market \cite{James2007LloydsOverview, Venezian1985RatemakingInsurance, Heinrich2021AHomogeneity, Zhou2013ApplicationCycles}. The first experiment, explores the interactions between syndicates and brokers with simplistic pricing methods to demonstrate a profitable market with low volatility. The second experiment incorporates catastrophe events that increase volatility in the market which can lead to insolvencies and pronounced cyclicality. Thirdly, the decision-making of syndicates is advanced with the introduction of improved exposure management, where syndicates become better equipped in dealing with catastrophe losses. Finally, we introduce the syndication of risk in the marketplace, via lead-follow mechanics, which significantly reduces volatility and tightly couples syndicates' loss experiences.

Some initial findings from the model have demonstrated stylised facts observed in the specialty insurance market, Lloyd's of London, thus validating the model. For example, when insurers price, based on past loss history with uniform risks and attritional losses only, we find that premium prices tend towards the fair analytical price but with substantial variance. When catastrophe events occur which affect multiple risks simultaneously, this leads to large industry losses, which cause step drops in syndicate capitals and increases in premium prices immediately afterwards. When syndicates are able to use exposure management processes to manage tail risks, this leads to smaller and better placed syndicate portfolios. Lastly, when lead and follow quote negotiation and line sizes are applied to premiums and claims, this improves the actuarial price estimates and reduces the spread between syndicates. Syndicate performance becomes more correlated as a result. 

The following sections of the article are Literature Review~\ref{literature_review-section} where pre-existing insurance based literature is disseminated and strengths/weaknesses highlighted. Model Description~\ref{model_description-section} describes every process, event and underlying functionality of the model. Results \& Discussion~\ref{results_discussion-section} introduces the specialty insurance market in the real-world, then delves into the aforementioned four experiment scenarios and discusses the observed quantitative and qualitative results. Lastly, the Conclusion~\ref{conclusion-section} highlights the findings from the model, how it contributes to specialty insurance research and future avenues to be explored.

\section{Literature review}
\label{literature_review-section}
The specialty insurance market differs from the conventional retail insurance market. The main difference is that the risks are intrinsically complex, offering cover for perils such as kidnap and ransom, cybersecurity breaches, terrorist attacks and political violence, commercial property damage and personal accident. These risks typically require specialist/expert assessment by insurers/syndicates \cite{DeMot2014SpecialLiability}. Simplifying the process extensively, a client reaches out to a broker with a risk, the broker brings the risk to the Lloyd's of London marketplace. Underwriters that operate within the Lloyd's market are reached out to, the underwriters utilise their subject matter expertise, actuarial pricing strategies and portfolio management approaches to decide if they should offer a lead, follow line or no line at all. If an underwriter offers a lead line, they underwrite a portion of the risk and tender a policy as a lead insurer, usually the lead covers a bigger portion of the risk and thus is paid a bigger share of the premiums, subsequently taking on more risk. Every risk is insured by a single lead insurer who usually shapes the policy and multiple follow insurers who agree to the terms and offer a line size. Once the risk has been covered proportionally across many syndicates, premiums are distributed depending on the line size agreed by all parties. Conversely, if a claim comes through, then everyone pays out proportionally, see \cite{Thoyts2010InsurancePractice, James2007LloydsOverview}.

The hardening and softening of the insurance market has been an area of academic enquiry over the years. The specialty insurance market can be a very costly, dynamic and highly volatile market to operate within, one characteristic of the market is the underwriting cycle. In short, when capacity is high, rates decrease and profits decrease, conversely, losses occur, insurers go insolvent or withdraw and capacity decreases, with subsequent upwards pressure on premium levels to generate acceptable return on investment. This is a typical soft to hard market transition, respectively \cite{boyer2012}, which attracts new capacity. The cycle then repeats.

The majority of the literature, has focused on this underwriting cycle phenomena, and have utilised agent-based, mathematical and time-series models to investigate the emergence and drivers of these cycles \cite{boyer2012, Owadally2019AnMarkets, Owadally2018TheCrises, Heinrich2021AHomogeneity, Weiss2004U.S.Capacity, Harrington1994PriceMarkets, Venezian1985RatemakingInsurance}. Surprisingly, there seems to be contradictory views with regard to the causes of market cycles, for example, \cite{Cummins1997PriceMarkets} argues that institutional lags are the main reason behind the liability insurance market cycle in the mid-1980s in North America when analysing insurance company level data. However, when analysing market-level data, they claim the opposite. Most in-depth literature reviews \cite{Boyer2015UnderwritingImagination, Harrington1994PriceMarkets, Weiss2004U.S.Capacity, boyer2012} agree that research into underwriting cycles intensified in the mid-1980s due to the ``liability crisis". During this time, an interesting discovery made by \cite{Venezian1985RatemakingInsurance} was that different lines of insurance have quasi-cyclical behaviour but others can't be discerned from white noise fluctuations, i.e., random walk. Due to these discoveries, Venezian et al. proposed auto-regressive (AR) models which replicated these cyclical behaviours. Following the findings from \cite{Venezian1985RatemakingInsurance}, most researchers primarily utilised AR models to investigate cyclicality in the insurance market. Furthermore, to ensure these models lead to actionable insights empirically, \cite{Cummins1997PriceMarkets} proposes that data used for analysing market cycles has traditionally been at the aggregate, industry level at a yearly temporal resolution, to remedy this, they argue for comprehensive datasets captured at a more granular spatio-temporal resolution. 

There are several hypothesised causes of underwriting cycles discerned by researchers, these are:

\begin{itemize}
    \item The flow of capital into and out of the market in response to market conditions such as major catastrophe events (capital shock) \cite{Venezian1985RatemakingInsurance, Winter1994TheMarkets}, while some disagree \cite{Choi2002TheModels}. 
    \item General economic conditions, i.e., rise and fall of central bank interest rates \cite{Choi2002TheModels, Grace1995ExternalCycle, Haley1995AIndustry}, contrary views \cite{Harrington2003DoRoots, Leng2006AnalysisInsurance}.
    \item Unanticipated inflation \cite{Venezian1985RatemakingInsurance, Karl2019HowMarkets}, while \cite{Leng2006AnalysisInsurance} disagrees.
    \item Institutional delays and lags, i.e., data collection, regulation and renewal periods \cite{Venezian1985RatemakingInsurance, Cummins1997PriceMarkets} while \cite{Cummins1997PriceMarkets} also makes a counter argument.
    \item Forecasting errors, imperfect knowledge of the market \cite{Venezian1985RatemakingInsurance, Cummins1997PriceMarkets, boyer2012}.
\end{itemize}

One of the prevailing reasons for why the analysis of market cycles has been inadequate is due to the cycle periods being longer than periods over which data is sampled \cite{Boyer2015UnderwritingImagination, boyer2012}. Furthermore, traditional time-series approaches work well in analysing deterministic systems, however, given the volatile nature of financial systems, this approach can be limited \cite{Ding1996ModelingApproach}. Moreover, if a modelled system can be affected by external factors then these factors should explicitly be included in the time-series model \cite{Hyndman2013CoherentModels}. For this and many other aforementioned reasons, this research article is well-timed and necessary.

Given the interest in modelling the insurance market, researchers have adopted individual-based modelling approaches, to capture new insights and overcome some of the earlier described issues \cite{Owadally2018TheCrises, Owadally2019AnMarkets, Zhou2013ApplicationCycles, England2022AnMouth, Dubbelboer2017AnInsurance, Heinrich2021AHomogeneity}. Some researchers have utilised individual-based models to discover stylised facts \cite{Giardina2003BubblesModels} such as the emergence of market crashes and bubbles. \cite{Owadally2019AnMarkets, boyer2012} argue that traditional methods such as AR models, fall short in quantifying the existence of a cycle in time-series data and questions the ability for these methods to forecast accurately. The article \cite{Owadally2019AnMarkets} suggests that linear models are insufficient and other approaches used to detect periodicity should be considered. Individual-based modelling approaches are more appealing as they can model the individual behaviours and social interactions at the micro level and generate complex aggregate behaviours such as cycles at the macro level \cite{Owadally2018TheCrises, Farmer2022Agent-BasedFuture}. Some researchers agree that the insurance market is complex, irrational and heterogeneous, evidenced by the interactions of individuals such as actuaries, underwriters, claims adjusters, brokers and organisations such as syndicates, regulators, managing agents and capital providers \cite{Ingram2010TheRationalities, Ingram2013CollectiveTheory, Fitzpatrick2004BarriersThem, Feldblum2001UnderwritingStrategies}. The model developed by \cite{Owadally2018TheCrises} tests the hypothesis that agents with simple decision-making rules interacting at the individual-level can produce aggregate complex behaviours depicting empirically valid market dynamics. \cite{Owadally2018TheCrises} also tests the theory of plural rationality \cite{Ingram2010TheRationalities, Ingram2013CollectiveTheory}, where they can parameterise the insurers mark-up calculation where a policy is either priced aggressively or priced using the actuarial fair price (market value). They found that as the pricing aggression increased, market volatility also increased due to large price fluctuations. 

Research conducted in \cite{Heinrich2021AHomogeneity} proposes an agent-based simulation of the specialty insurance market. The premise of the research is to investigate the regulatory changes under Solvency II that came into force in 2016. The model proposed, agents and processes represented as re-insurers, insurers, customers, shareholders and cat bonds that interact with each other and share information. The article finds that when the number of risk models increases for an insurer, i.e., an insurer utilises several catastrophe models (these are processes used to evaluate and manage natural and man-made catastrophe risk from perils, for example, hurricanes, wildfires) more risks are insured and conversely fewer risk models lead to lower profitability and higher risk of defaulting. The research article provides many interesting results with regards to catastrophe modelling approaches, to advanced exposure management features, whereby each insurer follows real-world regulations such as Solvency II and the ``Solvency Capital Requirement". This requirement is described as \textit{Insurers are required to have 99.5\% confidence they could cope with the worst expected losses over a year. That is, they should be able to survive any year-long interval of catastrophes with a recurrence frequency equal to or less than 200 years.} However, in the model of \cite{Heinrich2021AHomogeneity} some important aspects of the specialty insurance market have been neglected. These include attritional losses \cite{Owadally2018TheCrises, England2002FinancialInsurance} as well as the syndication of risk, a unique feature of Lloyd's of London \cite{Herschaft2004NotInsurance}, and finally the temporal granularity is fixed. Our proposed solution addresses the limitations of previous work. We believe our model is the first discrete event simulation of the specialty insurance market inspired by real-world actors within Lloyd's of London. 

A key insight and novelty of our article is in the use of a novel DES framework as opposed to the traditional ABM approach adopted by the aforementioned articles \cite{Owadally2019AnMarkets, Owadally2018TheCrises, Zhou2013ApplicationCycles, Farmer2009TheModelling, England2022AnMouth, Dubbelboer2017AnInsurance, Heinrich2021AHomogeneity}. Unlike the classical ABM approach, the DES approach captures the time-irregular and asynchronous nature of events in the specialty insurance market whereby events (such as insurance claims) occur at irregular time intervals from as little to as many events occurring simultaneously. Handling such unpredictable, irregular and asynchronous events in a classic ABM approach can be challenging, but is seamlessly handled using the DES approach. The benefits of DES have thoroughly been discussed in the following articles \cite{Hill2001ApplicationsProblems,Nevins1998AOperations,Jacobson2006Discrete-eventSystems}. Our model architecture is highly modular compared to the other approaches \cite{Owadally2018TheCrises, Owadally2019AnMarkets, Farmer2009TheModelling, Heinrich2021AHomogeneity}, as evidenced in the workflow diagram~\ref{fig:model_architecture_flow}, this allows the user to ``plug-in" with ease different pricing models (actuarial pricing, underwriter markup, exposure management), relationship networks (the relationships between syndicates and brokers), loss generators (attritional, catastrophic or both) and many output metrics such as yearly, monthly syndicate capitals, premiums offered and accepted, industry statistics (market health indicators), syndicate performance (syndicate health indicators), syndicate insolvency, catastrophe event (the risks affected and extent of losses), claims per risk and syndicates assigned to risk. These configuration processes and metrics will be described in detail in the Model Description~\ref{model_description-section} and Results \& Discussion~\ref{results_discussion-section} sections. Moreover, data analysis notebooks describing the stylised facts and unique features of the model will be provided as supplementary materials, some of these include, catastrophe losses and its impact on market health, advanced syndicates with exposure management modules and how they deal with losses, syndicate relationships with multiple brokers and its impact on risk coverage, capital venting strategies and what this means for the market. 

There are many reasons why insurers and/or researchers would utilise our proposed model to simulate the specialty insurance market dynamics. Practitioners may want a deeper understanding of the market dynamics, such as underwriting cycles, which drive the stability and profitability of the market. Moreover, the model provides both a quantitative and qualitative picture of the impact and drivers of these dynamics. Lastly, the model could be used to develop heuristics which enable better estimation of the current and future market conditions.

\section{Model description}
\label{model_description-section}
\begin{sidewaysfigure}[!htbp]
    \centering
    \includegraphics[width=0.90\textwidth]{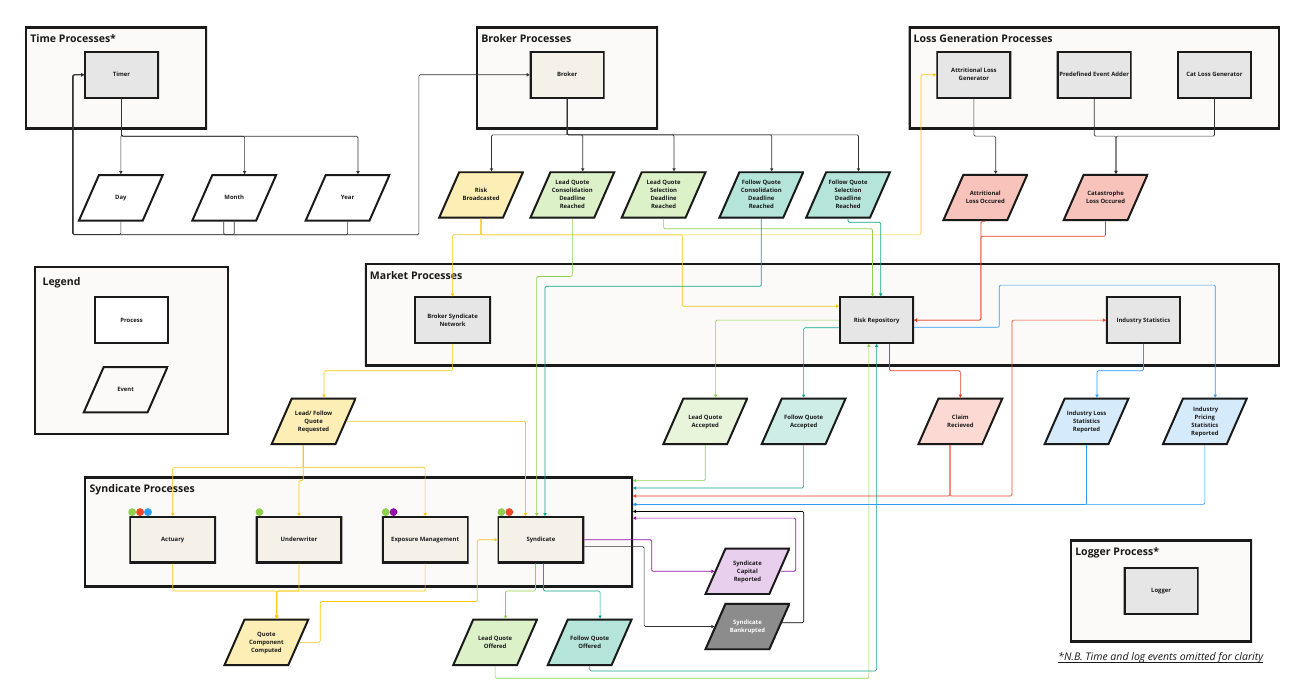}
    \caption{Workflow diagram of the specialty insurance market discrete event simulation. The line colours convey unique relationships between processes and events, i.e., yellow: new risk to quote requested and quote component generated, green: lead quote generation and selection, teal: follow quote generation and selection, red: losses to claims, blue: industry level statistics update, purple: syndicate level capital update.}
    \label{fig:model_architecture_flow}
\end{sidewaysfigure}

The purpose of the Lloyd's of London model is to investigate the emergent, complex characteristics which arise from individual-level interactions of the specialty insurance market. In this pursuit, the model conceptualises the behaviours of syndicates and brokers with which complex processes are undertaken as shown in Figure~\ref{fig:model_architecture_flow}. These individual-level interactions from the bottom-up lead to the emergence of stylised facts at the aggregate level which correspond to empirical market trends. In this section, we describe the main agents/processes which the model encapsulates, the events they respond to and emit (i.e. the actions they undertake) and how these features all fit together to produce the DES. We start this section by discussing the notion of time in the model and its importance.

\subsection{Time}
Discrete Event Simulations do not necessarily require the notion of time compared to other methods such as ABMs. Instead, events trigger processes which the process in turn responds to by emitting additional events and/or performing actions, i.e., updating an internal state variable. In theory, there is no need for the concept of time. However, in reality, there is a notion of time associated with events. For simulations of complex financial systems, such as the specialty insurance market, the challenge is that events occur asynchronously while the timescale is irregular and varying. Capturing these features of the specialty insurance market is both paramount (in order to ensure the model reflects empirical trends) and difficult to achieve with a typical ABM as mentioned in the following \cite{Zhou2013ApplicationCycles, Heinrich2021AHomogeneity, Owadally2019AnMarkets}. Conversely, the flexibility of the DES framework, allows events to have an associated timestep, which only triggers once the timestep in question occurs, e.g., $t = 3$. In our DES, a regular timer is incorporated with a standard timesteps of days, months and years. This allows us to capture events which occur in the simulation at different levels of granularity. Moreover, we are able to investigate different emergent phenomena independent of the timescale across which they occur. 

The regular timer in our model generates the events shown in Table~\ref{tab:time_events}. We also detail the processes which respond to the events and the description of the events.

\begin{table}[!htbp]
\centering
\resizebox{10cm}{!}{\begin{tabular}{|p{1cm}|p{4cm}|p{5cm}|}
\hline
\textbf{Event} & \textbf{Processes which respond to event} & \textbf{Description} \\\hline
Day & Broker Process & The smallest unit of time in the model.  Trigger processes that must respond on a daily basis.  \\\hline
Month & Syndicate Process, Actuarial Sub-Process, Central Risk Repository Process  & Trigger processes that must respond on a monthly basis. \\\hline
Year & Syndicate Process, Actuarial Sub-Process, Underwriting Sub-Process, Premium Exposure Management Sub-Process, Value at Risk (VaR) Exposure Management Sub-Process & Trigger processes that must respond on a yearly basis.  \\\hline
\end{tabular}}
\caption{Events which the time process generates.}
\label{tab:time_events}
\end{table}

\subsection{Broker process}

The primary purpose of the Broker process is to bring new risks to the market. This occurs in response to a Day event as noted in Table~\ref{tab:time_events}. Upon the triggering of a Day event, the broker generates $n_{r}$ risks according to a Poisson distribution where the $\lambda$ variable for the distribution is given by the risks per day (RPD) input parameter as described in Table~\ref{tab:input_parameters}. The Broker process emits a number of events when the risk is generated in order to broadcast the risk to the insurers/syndicates and to set deadlines for the quotes from the insurers to be finalised.

The broker responds to and generates the events shown in Table~\ref{tab:broker_events_respond_generate}

\begin{table}[H]
\centering
\large
\resizebox{10cm}{!}{\begin{tabular}{|p{3cm}|p{4cm}|p{4cm}|p{5cm}|}
\hline
\textbf{Event} & \textbf{Processes which respond to event}  &\textbf{Processes which generate event} & \textbf{Description} \\\hline
Day & Broker Process & Time Process & Triggers the broker process to generate new risks.  \\\hline
Risk Broadcasted & Broker-Syndicate Network Process, Central Risk Repository Process, Attritional Loss Generator Process, Industry Statistics Process  & Broker Process &  Broadcasts a new risk has entered the market. The responding processes perform certain actions in response.  \\\hline
Lead Quote Consolidation Deadline Reached & Syndicate Process & Broker Process & Sets a deadline for any lead syndicates who plan on sending an offer to consolidate and send the quote offer. \\\hline
Lead Quote Selection Deadline Reached & Central Risk Repository Process & Broker Process & Sets a deadline for a lead syndicate to be chosen based on the minimum quote offered by the syndicates.  \\\hline
Follow Quote Consolidation Deadline Reached & Syndicate Process & Broker Process & Sets a deadline for any follow syndicates who plan on sending an offer to consolidate and send the quote offer  \\\hline
Follow Quote Selection Deadline Reached & Central Risk Repository Process & Broker Process & Sets a deadline for follow syndicates to be chosen based on the line sizes offered by the syndicates. \\\hline
\end{tabular}}
\caption{Events which the broker process responds to and generates.}
\label{tab:broker_events_respond_generate}
\end{table}

\subsection{Broker-syndicate network process}

The broker-syndicate network process is responsible for requesting quotes for new risks entering the market from the registered syndicates. This is facilitated by a network topology which for a given risk, selects a number of syndicates to which a quote is requested. Therefore, a risk does not necessarily have to be broadcasted to all syndicates depending on the parameter choices. The options for the topologies are the following:

\begin{itemize}
    \item Circular topology: inspired by the model presented in \cite{Owadally2018TheCrises}.
    \item Network topology: inspired by the interviews and workshops with stakeholders at Ki \& Brit Syndicates.
    \item Random topology: is a base case feature.
\end{itemize}

Prevalent in all the topologies is a lead and follow $top\_k$, parameter as mentioned in Table~\ref{tab:input_parameters}. This parameter selects the best $k$ syndicates based on the chosen topology methods. For instance, in the circular topology \cite{Owadally2018TheCrises}, the distance between the brokers and syndicates is used as a measure of the ``ease of doing business" where a small distance implies a low cost of doing business conversely a large distance implies a larger cost. The syndicates are then ordered based on the $top\_k$ parameter which selects the $k$ lowest cost syndicates. The network topology represents a connected network/graph between the brokers and syndicates where the edge weightings represent the ease of doing business between the brokers and syndicates. The larger the edge weighting, the more likely for brokers and syndicates to do business and vice-versa. Once again, the syndicates are ordered based on the strongest relationship with the $top\_k$ parameter filtering this down to the strongest $k$ relationships. Lastly, in the random topology (the adopted network for the experiments conducted below), as the name suggests, the syndicates are randomly ordered with the $top\_k$ parameter selecting the first $k$ syndicates in the list. 

Given the above, the broker-syndicate network responds to and generates the events shown in Table~\ref{tab:broker_syndicate_network_events_respond_generate}.

\begin{table}[H]
\centering
\large
\resizebox{10cm}{!}{\begin{tabular}{|p{2cm}|p{4cm}|p{4cm}|p{5cm}|}
\hline
\textbf{Event} & \textbf{Processes which respond to event} & \textbf{Processes which generate event} & \textbf{Description} \\\hline
Risk Broadcasted & Broker-Syndicate Network Process & Broker Process & Triggers broker-syndicate network to select $k$ syndicates to request a Leader Quote and $k$ syndicates for a Follower Quote. \\\hline
Lead Quote Requested & Syndicate Process & Broker-Syndicate Network Process &  Triggers syndicate process to calculate a price and lead line size for the risk if possible.  \\\hline
Follow Quote Requested & Syndicate Process & Broker-Syndicate Network Process &  Triggers syndicate process to calculate a follow line size based on leaders terms if possible.   \\\hline
\end{tabular}}
\caption{Events which the broker-syndicate network process responds to and generates.}
\label{tab:broker_syndicate_network_events_respond_generate}
\end{table}

\subsection{Central risk repository process}

The central risk repository process is responsible for tracking all of the risks, quotes and underwritten policies in the market. This includes all of the quotes offered for a risk, the policy information if the risk has been underwritten such as who the leader and follower syndicates are. Finally, the central risk repository is responsible for applying any losses, whether attritional or catastrophic, to the underlying syndicates on the policy in question. 

Given the above, the central risk repository responds to and generates the events shown in Table~\ref{tab:crr_events_responded_generated}.

\begin{table}[H]
\centering
\large
\resizebox{10cm}{!}{\begin{tabular}{|p{3cm}|p{4cm}|p{4cm}|p{5cm}|}
\hline
\textbf{Event} & \textbf{Processes which respond to event} & \textbf{Processes which generate event} & \textbf{Description} \\\hline
Month & Central Risk Repository Process & Time Process & Triggers central risk repository process to calculate monthly statistics about the risks, quotes and policies. \\\hline
Risk Broadcasted & Central Risk Repository Process & Broker Process & Registers new risk entering the market. \\\hline
Lead Quote Offered & Central Risk Repository Process & Syndicate Process & Registers a lead quote offered by a syndicate for a given risk. \\\hline
Lead Quote Selection Deadline Reached & Central Risk Repository Process & Broker Process & Triggers the deadline by which no more lead quote offers are accepted. The central risk repository process selects the leader based on the cheapest quote.\\\hline
Follow Quote Offered & Central Risk Repository Process &  Syndicate Process & Registers a follow quote offered by a syndicate for a given risk.  \\\hline
Follow Quote Selection Deadline Reached & Central Risk Repository Process & Broker Process & Triggers the deadline by which no more follow quote offers are accepted. The central risk repository checks to ensure that a leader has already been selected, how much line size remains and assigns the requested line size to the followers as long as it does not exceed 100\% whereby they are proportionally signed down.\\\hline
Attritional Loss Occurred & Central Risk Repository Process & Attritional Loss Generator Process & Triggers the attritional loss to be broadcasted to the syndicates on the policy. \\\hline
Catastrophe Loss Occurred & Central Risk Repository Process & Catastrophe Loss Generator Process & Triggers the catastrophe loss to be broadcasted to the syndicates on the policy. \\\hline
Lead Quote Accepted & Syndicate Process & Central Risk Repository Process & Notifies syndicate that their lead quote was accepted. \\\hline
Follow Quote Accepted & Syndicate Process & Central Risk Repository Process & Notifies syndicate that their follow quote was accepted. \\\hline
Claim Received & Syndicate Process & Central Risk Repository Process & Notifies syndicate that a claim was received for a risk they underwrite.\\\hline
Industry Pricing Statistics Reported & Industry Statistics Process & Central Risk Repository Process & Passes industry statistics for the risks, quotes and policies in the central risk repository. \\\hline
\end{tabular}}
\caption{Events which the central risk repository process responds to and generates.}
\label{tab:crr_events_responded_generated}
\end{table}

\subsection{Syndicate process}

The syndicate process represents one of the most detailed agents within the model, as it is responsible for a number of important functions. Primarily, the syndicate is responsible for pricing any risks which come to market and decide which line size to give. This is all done in the context of a capital management framework whereby syndicates must ensure they are appropriately capitalised even in the case of tail loss events occurring, in order to avoid insolvency. Lastly, the syndicate must also provide any dividend back to capital providers in the case of profitable performance. As the syndicate is responsible for a number of functions, this section has been split into a number of sub-sections according to the sub-processes which compose the main syndicate process. 

Before moving on to the sub-processes, we note that the main syndicate process is responsible for coordinating and organising the sub-processes which compose it. In this regard, the syndicate process responds to and generates the events shown in Table~\ref{tab:syndicate_events_responded_generated}.

\begin{table}[H]
\centering
\large
\resizebox{10cm}{!}{\begin{tabular}{|p{4cm}|p{6cm}|p{6cm}|p{6cm}|}
\hline
\textbf{Event} & \textbf{Processes which respond to event} &\textbf{Processes which generate event} & \textbf{Description} \\\hline
Month & Syndicate Process & Time Process & Triggers the syndicate process to share state variables, i.e., capital with its sub-processes. \\\hline 
Year & Syndicate Process & Time Process & Triggers the dividend sub-module to evaluate the profitability of the syndicate and pay a dividend if profitable. \\\hline
Lead Quote Requested & Syndicate Process & Broker-Syndicate Network Process & Triggers the syndicate process to note that a lead quote has been requested so that the actuarial, underwriting, line size and exposure management sub-processes can prepare or refuse a quote. \\\hline
Follow Quote Requested & Syndicate Process & Broker-Syndicate Network Process & Triggers the syndicate process to note that a follow quote has been requested so that the actuarial, underwriting, line size and exposure management sub-processes can prepare or refuse a quote. \\\hline
Quote Component Computed & Syndicate Process & Actuarial Sub-Process, Underwriting Sub-Process, Premium Exposure Management Sub-Process, VaR Exposure Management Sub-Process & Triggers the syndicate process to take the different components of the quote and store all of these components until a quote consolidation deadline is reached. \\\hline
Lead Quote Consolidation Deadline Reached & Syndicate Process & Broker Process & Triggers the syndicate to take all the stored quote components of a lead quote and to either emit a LeadQuoteOffered event or if one of the components rejects the quote (i.e., exposure management deems it risky) then nothing is emitted. \\\hline
Lead Quote Accepted & Syndicate Process & Central Risk Repository Process & Notifies the syndicate that their lead quote has been accepted by the broker and they now underwrite the policy (i.e., responsible for claims and collect premiums). \\\hline
Follow Quote Consolidation Deadline Reached & Syndicate Process & Broker Process & Triggers the syndicate to take all the stored quote components of a follow quote and to either emit a FollowQuoteOffered event or if one of the components rejects the quote (i.e., exposure management deems it risky) then nothing is emitted. \\\hline
Follow Quote Accepted & Syndicate Process & Central Risk Repository Process & Notifies the syndicate that their follow quote has been accepted by the broker and they now underwrite the policy (i.e. responsible for claims and collect premiums). \\\hline
Claim Received & Syndicate Process & Central Risk Repository Process & Notifies the syndicate that a claim has been received for one of the policies they underwrite. The claim is paid and in case of insolvency, the syndicate emits a SyndicateBankrupted event.\\\hline

Lead Quote Offered & Central Risk Repository Process & Syndicate Process & Triggers the central risk repository to note that a lead quote is offered by the syndicate for the risk in question. \\\hline
Follow Quote Offered & Central Risk Repository Process & Syndicate Process & Triggers the central risk repository to note that a follow quote is offered by the syndicate for the risk in question. \\\hline
Syndicate Bankrupted & Actuarial Sub-Process, Underwriting Sub-Process, Premium Exposure Management Sub-Process VaR Exposure Management Sub-Process, Line Size Sub-Module, Dividend Sub-Module & Syndicate Process & Triggers the DES framework to remove the syndicate and all of its sub-processes from the simulation. \\\hline
Syndicate Capital Reported & Actuarial Sub-Process, Underwriting Sub-Process, Premium Exposure Management Sub-Process VaR Exposure Management Sub-Process, Line Size Sub-Module, Dividend Sub-Module & Syndicate Process & Updates the shared state of capital among the syndicate sub-processes. \\\hline
\end{tabular}}
\caption{Events which the syndicate process responds to and generates.}
\label{tab:syndicate_events_responded_generated}
\end{table}

\subsubsection{Actuarial sub-process}
\label{sub-section:actuarial}
Pricing of risks in the insurance market is a complex process, risks are heterogeneous, extremely variable and fundamentally stochastic and ambiguous \cite{Barrett1999ChallengesMarket, Dionne2000HandbookInsurance, Heinrich2021AHomogeneity, Owadally2019AnMarkets}. Our model captures all of these features but makes the simplifying assumption that all risks are homogeneous and that they belong to a catastrophe exposed class such as property insurance. The price of a risk will depend on several factors, primarily, the nature of the risk itself (i.e. the frequency and severity with which the risk can occur) and the market's view of the risk (i.e. the supply and demand of services in the market) \cite{Owadally2019AnMarkets, Zhou2013ApplicationCycles, Kunreuther1989TheRisks}.

These two factors are different from one another. The first, is related to the quantification of the risk itself without considering any external market influences. This process is often carried out by an actuary \cite{Zhou2013ApplicationCycles, Owadally2018TheCrises} The actuary will assess the risk on experience based metrics (historical data on the risk or similar risks) and/or exposure based metrics (quantification of the risk in absence of data but based on a risk profile). Fundamentally, the actuary attempts to come up with a ``fair price" for the risk based on the expectation of losses. That is to say, a price which would cover the expected losses of the risk over its lifetime. The second factor, considers the price of the risk based on the market's view i.e. the supply and demand of insurance services. This process is typically carried out by an underwriter. The underwriter, will take the fair price guideline from the actuary and, in very simplistic terms, scale this price up or down based on the supply and demand in the market. For instance, although the actuary may propose a much higher fair price, the market trends may force the underwriter to reduce this price in order to be competitive. 

In this section, we will focus on the actuarial sub-process, inspired by the work of \cite{Owadally2019AnMarkets, Zhou2013ApplicationCycles}. In the next section, we will discuss the underwriting sub-process. 

Based on the above, the actuarial sub-process price is given by two main components. The first component is the insurer's expected claim cost:

\begin{equation}
\label{eq:insurers_expected_claim_cost}
    \Tilde{P}_t = z\bar{X}_t + (1-z)\lambda'_t\mu'_t
\end{equation}

where $\Tilde{P}_t$ is the insurers expected claim cost at timestep $t$, $\bar{X}_t$ is the insurers past weighted average claims, $\lambda'_t$ is the industry-wide average claim frequency, $\mu'_t$ is the industry-wide expected claim cost and finally $z$ is the internal experience weight input parameter Table~\ref{tab:input_parameters} which decides whether a syndicate weighs their own loss or the industry loss experience as more important. As per \cite{Owadally2019AnMarkets, Zhou2013ApplicationCycles}, the insurers expected claim cost, $\bar{X}_t$, is calculated as a simple exponentially weighted moving average where the weight is an input parameter to the model Table~\ref{tab:input_parameters} called the loss experience recency weight. 

The final actuarial price, which we denote as $P_{at}$, is the sum of the insurers expected claim cost, $\Tilde{P}_t$, and a ``risk loading term", $\alpha F_t$, where $F_t$, is the standard deviation of the insurer's claims while $\alpha$ is a input parameter to the model Table~\ref{tab:input_parameters}, called the volatility weight:

\begin{equation}
\label{eq:actuarial_price}
    P_{at} = \Tilde{P}_t + \alpha F_t
\end{equation}

As can be seen, the actuarial price in Equation \ref{eq:actuarial_price}, captures the main idea of syndicates pricing a risk to cover their expected claim losses as well as to allow for some volatility in the losses. 

The actuarial sub-process responds to and generates the events shown in Table~\ref{tab:actuarial_events_respond_generate}.

\begin{table}[H]
\centering
\large
\resizebox{10cm}{!}{\begin{tabular}{|p{3cm}|p{5cm}|p{5cm}|p{5cm}|}
\hline
\textbf{Event} & \textbf{Processes which respond to event} & \textbf{Processes which generate event} & \textbf{Description} \\\hline
Quote Requested & Actuarial Sub-Process & Syndicate Process & Triggers actuarial sub-process to calculate the actuarial price as per Equations~\ref{eq:insurers_expected_claim_cost} and \ref{eq:actuarial_price}. \\\hline
Industry Loss Statistics Reported & Actuarial Sub-Process & Industry Statistics Process & Distributes the latest industry statistics so that the actuarial sub-process can use this in Equation~\ref{eq:insurers_expected_claim_cost}.\\\hline 
Industry Pricing Statistics Reported & Actuarial Sub-Process & Industry Statistics Process & Distributes the latest industry pricing statistics so that the actuarial sub-process can use this in Equation~\ref{eq:insurers_expected_claim_cost}.\\\hline 
Claim Received & Actuarial Sub-Process & Central Risk Repository Process & Triggers the actuarial sub-process to update the syndicate's claim experience so it can be used in Equation~\ref{eq:insurers_expected_claim_cost}. \\\hline 
Syndicate Bankrupted & Actuarial Sub-Process & Syndicate Process & Triggers the actuarial sub-process to be removed from the simulation after its associated syndicate becomes insolvent. \\\hline 
Quote Component Computed & Syndicate Process & Actuarial Sub-Process & Sends the associated syndicate the actuarial pricing component i.e. Equation~\ref{eq:actuarial_price}.  \\\hline
\end{tabular}}
\caption{Events which the actuarial sub-process responds to and generate.}
\label{tab:actuarial_events_respond_generate}
\end{table}

\subsubsection{Underwriting sub-process}

As explained in the previous section, the objective of the underwriting sub-process is to ``scale" the actuarial price in order to match market supply and demand. We again, employ the equations used by \cite{Owadally2019AnMarkets, Zhou2013ApplicationCycles}, where they apply neoclassic price theory for the price-elasticity of demand. The details of the derivation can be found in \cite{Owadally2019AnMarkets, Zhou2013ApplicationCycles}. The underwriter scaling is given by:

\begin{equation}
\label{eq:uw_markup}
    P_{t} = P_{at}e^{m_t}
\end{equation}

where $P_{t}$ is the final price offered by the syndicate after the underwriters scaling, $P_{at}$ is the actuarial price from Equation~\ref{eq:actuarial_price} and $m_t$ is the underwriter log markup which attempts to model the price-elasticity of demand in the market. 

The underwriter markup, $m_t$, is also calculated as a simple exponentially weighted moving average where the weight is an input parameter to the model Table~\ref{tab:input_parameters} called the underwriter markup recency weighting. Details on how the underwriter markup is calculated can be found in \cite{Owadally2019AnMarkets, Zhou2013ApplicationCycles}.

The underwriting sub-process responds to and generates the events shown in Table~\ref{tab:underwriting_events_respond_generate}.

\begin{table}[H]
\centering
\large
\resizebox{10cm}{!}{\begin{tabular}{|p{3cm}|p{5cm}|p{5cm}|p{5cm}|}
\hline
\textbf{Event} & \textbf{Processes which respond event} & \textbf{Processes which generate event} & \textbf{Description} \\\hline
Quote Requested & Syndicate Process & Underwriting Sub-Process & Triggers underwriting sub-process to calculate the final price as per Equations~\ref{eq:uw_markup}. \\\hline
Syndicate Bankrupted & Syndicate Process & Underwriting Sub-Process & Triggers the underwriting sub-process to be removed from the simulation after its associated syndicate becomes insolvent. \\\hline 
Quote Component Computed & Underwriting Sub-Process & Syndicate Process & Sends the associated syndicate the underwriting pricing component i.e. Equation~\ref{eq:uw_markup}. \\\hline
\end{tabular}}
\caption{Events which the underwriting sub-process responds to and generate.}
\label{tab:underwriting_events_respond_generate}
\end{table}

\subsubsection{Value at Risk (VaR) exposure management sub-process}

Exposure management is a critical function for all insurance firms, to such that appropriate levels of exposure management are regulated by law for insurers. Exposure management allows insurers to have a quantitative understanding of the tail risk associated with the policies that they underwrite. By doing so, insurers can quantify the impact of worst case scenarios on their portfolio. This informs insurer's underwriting strategy. For instance, given the state in Figure \ref{fig:catastrophic_events_peril}, an over-exposed strategy would be for lead syndicate B to underwrite all risks within region 1. Therefore, exposure management can alert syndicates if they are over-exposed to a given catastrophe peril region.

The Value at Risk (VaR) of an insurer's portfolio is a common measure used to quantify the level of risk taken by an insurer. The VaR with some exceedance probability, which we denote as $\alpha$ Table~\ref{Input_Params}, identifies the amount of syndicate capital which is at risk in case of any tail events occurring e.g. exceedingly large catastrophe events. The exposure management sub-process is therefore responsible for ensuring that the syndicate capital remains above this threshold value. Given its importance, our model employs the VaR exposure management detailed in \cite{Heinrich2021AHomogeneity}.  

The VaR Exposure Management sub-process responds to and generates the events shown in Table~\ref{tab:var_events_respond_generate}.

\begin{table}[H]
\centering
\large
\resizebox{10cm}{!}{\begin{tabular}{|p{3cm}|p{5cm}|p{5cm}|p{5cm}|}
\hline
\textbf{Event} & \textbf{Processes which respond to event} & \textbf{Processes which generate event} & \textbf{Description} \\\hline
Simulation Started & VaR Exposure Management Sub-Process & Predefined Event\tablefootnote{A special event initialised by the DES framework.} & Triggers the VaR EM sub-process to perform the VaR simulations for the various peril regions. \\\hline
Quote Requested & VaR Exposure Management Sub-Process & Syndicate Process & Triggers the VaR EM sub-process to check its exposure management to the peril region in question. If the risk can not be underwritten then the QuoteComponentComputed event reflects this. \\\hline
Syndicate Capital Reported & VaR Exposure Management Sub-Process & Syndicate Process & Updates the common state of capital between the syndicate process and the EM sub-process. \\\hline
Syndicate Bankrupted & VaR Exposure Management Sub-Process & Syndicate Process & Triggers the VaR EM sub-process to be removed from the simulation after it’s associated syndicate becomes insolvent. \\\hline
Quote Component Computed & Syndicate Process & VaR Exposure Management Sub-Process & Sends the associated syndicate the VaR EM quote component. This either allows the quote to go ahead or vetoes the quote in case the capital requirements of the syndicate do not satisfy the minimum exposure management standards.\\\hline
\end{tabular}}
\caption{Events which the VaR EM sub-process responds to and generates.}
\label{tab:var_events_respond_generate}
\end{table}

\subsubsection{Premium exposure management sub-process}

In the previous section, one approach to exposure management was explored via the VaR measure. However, as extensively discussed by \cite{Heinrich2021AHomogeneity}, VaR exposure management is a complicated process, often relying on computationally expensive monte carlo simulations and in many cases, difficult-to-measure variables. Often insurers seek approximations or proxy approaches which capture the essence of the VaR exposure management. In this section, we detail such a methodology which we refer to as Premium Exposure Management.

The premium an insurer collects for underwriting a risk can be thought of as a proxy for exposure. In particular, when the total premium written is compared to the total capital available, this gives a proxy measure of how exposed an insurer is to the potential risk of insolvency. For example, if an insurer has underwritten a large number of risks, and therefore is collecting a large premium, but their total capital is comparatively smaller. This indicates that the insurer might have insufficient capital available to cover the full range of potential losses. On the other hand, if premiums written were large but the capital available was also large then this would indicate that the insurer has suffered minimal losses and that their current underwriting strategy was profitable. This is the essence of the premium exposure management sub-process. 

The Premium Exposure Management sub-process responds to and generates the events shown in Table~\ref{tab:premium_events_respond_generate}.

\begin{table}[H]
\centering
\large
\resizebox{10cm}{!}{\begin{tabular}{|p{3cm}|p{5cm}|p{5cm}|p{5cm}|}
\hline
\textbf{Event} & \textbf{Processes which respond to event} & \textbf{Processes which generate event} & \textbf{Description} \\\hline
Quote Requested & Premium Exposure Management Sub-Process & Syndicate Process & Triggers the Premium EM sub-process to calculate the premium EM scaling factor which is used to scale the quote price based on the current exposure. \\\hline
Syndicate Capital Reported & Premium Exposure Management Sub-Process & Syndicate Process & Updates the common state of capital between the syndicate process and the EM sub-process. \\\hline
Syndicate Bankrupted & Premium Exposure Management Sub-Process & Syndicate Process & Triggers the Premium EM sub-process to be removed from the simulation after it’s associated syndicate becomes insolvent. \\\hline
Quote Component Computed & Syndicate Process & Premium Exposure Management Sub-Process & Sends the associated syndicate the Premium EM quote component. This component is the scaling factor applied to the quote price based on the capital requirements of the syndicate.\\\hline
\end{tabular}}
\caption{Events which the Premium EM sub-process responds to and generates.}
\label{tab:premium_events_respond_generate}
\end{table}

\subsubsection{Line size sub-module}

A unique feature of the Lloyd's of London specialty insurance market is the syndication of risk i.e. there will typically be several syndicates on a policy which includes one lead syndicate and several follow syndicates. For this reason, the various syndicates will underwrite only a fraction of the risk, this is termed the line size. In our model, the lead syndicates offer a default line size and set the price of the policy based on the actuarial and underwriting sub-processes. The candidate follow syndicates, also price the risk based on their actuarial and underwriting sub-processes which they then compare to the actual price from the lead syndicate. This allows them to assess the ``pricing strength" which in turn is used to decide the line size to offer. The pricing strength is defined as the ratio of the follower's proposed price to the lead price. That is, if the pricing strength is above one, this implies the price of the risk is good and a larger line size is offered and vice versa. 

\subsubsection{Dividend sub-module}
As explored in \cite{Heinrich2021AHomogeneity}, our DES model also includes, the ability for syndicates to pay a dividend to capital holders on a yearly basis provided they have made a profit. The reason for why this is a class object and not a DES process, is because the only process that relies on its outputs is the syndicate process.

The dividend sub-module only becomes active when a Year event triggers the syndicate process. When this occurs, the syndicate uses the dividend sub-module to check whether a profit has been made and if so calculates the dividend as follows:

\begin{equation}
\label{eq:dividend}
    D = \gamma Pr_t
\end{equation}

where $D$ is the dividend paid, $Pr_t$ is the profit made by the syndicate, and $\gamma$, profit fraction, is an input parameter to the model Table~\ref{tab:input_parameters}, which represents the fraction of the profit to pay out as a dividend.

\subsection{Attritional loss generator process}
Attritional losses (as opposed to catastrophe losses) are defined as those losses which are generally uncorrelated with each other in both space and time, have high frequency, low severity and are fairly predictable \cite{England2002FinancialInsurance}. On the other hand, catastrophe losses, which will be discussed in the next sub-section, tend to be spatially correlated (affecting a number of policies concentrated by a given peril region, class or industry), low frequency, high severity and difficult to predict. 

In our model, we develop the attritional loss generator process inspired by the work of \cite{Owadally2018TheCrises, Owadally2019AnMarkets, Zhou2013ApplicationCycles}. The attritional loss generator process pre-generates a number of AttritionalLossOccurred events when the risk is first brought to the market. The number of claim events is given by the Poisson distribution with the $\lambda$ value set as the yearly claim frequency, an input parameter as defined in Table~\ref{tab:input_parameters}. The severity of the loss is defined by the gamma distribution. The shape parameter of the distribution is given by $\frac{1}{COV^2}$, where $COV$ is the gamma distribution's coefficient of variation as defined in Table~\ref{tab:input_parameters}. The scale parameter of the distribution is given by $\mu COV^2$ where $\mu$ is the mean of the gamma distribution. These AttritionalLossOccurred events are then placed within the event queue within the expiration date of the risk. 

Given the above, the attritional loss generator responds to and generates the events shown in Table~\ref{tab:attritional_loss_events_respond_generate}.

\begin{table}[H]
\centering
\resizebox{10cm}{!}{\begin{tabular}{|p{2cm}|p{4cm}|p{4cm}|p{5cm}|}
\hline
\textbf{Event} & \textbf{Processes which respond to event} & \textbf{Processes which generate event} & \textbf{Description} \\\hline
Risk Broadcasted & Attritional Loss Generator Process & Broker Process & Triggers the attritional loss generator process to pre-generate a number of AttritionalLossOccurred events as specified above. \\\hline
Attritional Loss Occurred & Central Risk Repository & Attritional Loss Generator Process & Triggers the central risk repository to broadcast the attritional loss to the syndicates on the policy in question.  \\\hline
\end{tabular}}
\caption{Events which the attritional loss generator process responds to and generates.}
\label{tab:attritional_loss_events_respond_generate}
\end{table}

\subsection{Catastrophe loss generator process}
\begin{figure}[!htbp]
    \centering
    \includegraphics[width=0.70\textwidth]{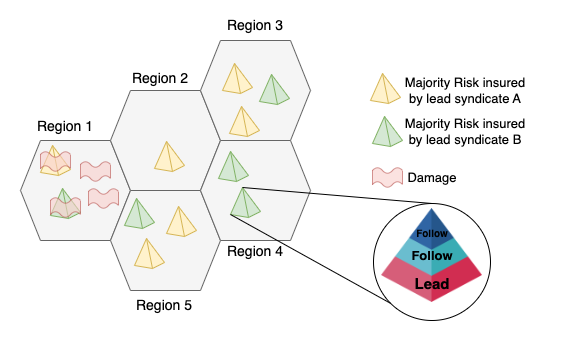}
    \caption{All risks are assigned to a given peril region, catastrophes affect a given peril region, in this case Region 1. The impact of the losses cascade down to each risk and the lead and follow syndicates on the policy affected.}
    \label{fig:catastrophic_events_peril}
\end{figure}

In our model, we develop the catastrophe loss generator process inspired by \cite{Heinrich2021AHomogeneity}. The catastrophe loss generator attempts to capture the phenomena that catastrophe losses are both correlated spatially (e.g. a number of policies concentrated by a given peril region) and temporally. For this reason, all risks in the model have an associated peril region, as observed in Figure~\ref{fig:catastrophic_events_peril}. Unlike the attritional loss generator, which generates attritional events and losses on a risk by risk basis, the catastrophe loss generator generates catastrophe events and losses on a peril region basis. The total loss affecting a given peril region is then cascaded down to the affected risks within the peril region, and subsequently the lead and follow insurers. 

When the simulation starts, the catastrophe loss generator pre-generates a number of catastrophe events over the length of the entire simulation. This is done via the Poisson distribution with the $\lambda$ value set as the product of the mean number of catastrophe events per year Table~\ref{tab:input_parameters} and the number of years in the simulation. A peril region is randomly assigned to the CatastropheLossOccurred event. The total loss affecting the peril region is given by a truncated Pareto distribution with the minimum value set as the minimum catastrophe damage Table~\ref{tab:input_parameters}. The CatastropheLossOccurred events are then added within the event queue up until the end of the simulation. 

Given the above, the catastrophe loss generator responds to and generates the events shown in Table~\ref{tab:catastrophe_loss_events_respond_generate}.

\begin{table}[H]
\centering
\large
\resizebox{10cm}{!}{\begin{tabular}{|p{3cm}|p{4cm}|p{4cm}|p{5cm}|}
\hline
\textbf{Event} & \textbf{Processes which respond to event} & \textbf{Processes which generate event} & \textbf{Description} \\\hline
SimulationStarted & Catastrophe Loss Generator Process & Pre-defined Event & Triggers the catastrophe loss generator process to pre-generate CatastropheLossOccurred events for the different peril regions as described above and shown in Figure~\ref{fig:catastrophic_events_peril}.. \\\hline
Catastrophe Loss Occurred & Central Risk Repository & Catastrophe Loss Generator Process & Triggers the central risk repository to cascade the total loss imposed on the peril region to the specific policies within the region. \\\hline
\end{tabular}}
\caption{Events which the catastrophe loss generator process responds to and generates.}
\label{tab:catastrophe_loss_events_respond_generate}
\end{table}

\subsection{Industry statistics process}

The industry statistics process, as the name suggests, simply keeps a track of all the relevant industry statistics and metrics. This is necessary, as many of the syndicate sub-processes rely on market-wide metrics. Note, that this process does not attempt to mimic any real market agents/processes. It is simply just an aggregator of data.  

The industry statistics process responds to and generates the events shown in Table~\ref{tab:industry_statistics_events_respond_generate}.

\begin{table}[H]
\centering
\large
\resizebox{10cm}{!}{\begin{tabular}{|p{3cm}|p{4cm}|p{4cm}|p{5cm}|}
\hline
\textbf{Event} & \textbf{Processes which respond to event} & \textbf{Processes which generate event} & \textbf{Description} \\\hline
Year & Industry Statistics Process & Time & Triggers the industry statistics process to send out yearly market-wide statistics.  \\\hline
Risk Broadcasted & Industry Statistics Process & Broker & Triggers the industry statistics process to log new risks entering the market.  \\\hline
Lead Quote Accepted & Industry Statistics Process & Central Risk Repository & Triggers the industry statistics process to log newly underwritten risk.  \\\hline
Claim Received & Industry Statistics Process & Central Risk Repository & Triggers the industry statistics process to log new claim events.  \\\hline
Industry Loss Statistics Reported & Actuarial Sub-Process & Industry Statistics Process & Sends industry loss statistics to the actuarial sub-process in order to perform relevant calculations. \\\hline
\end{tabular}}
\caption{Events which the industry statistics process responds to and generates.}
\label{tab:industry_statistics_events_respond_generate}
\end{table}

\begin{table}[!htbp]
\centering
\small
\resizebox{12cm}{!}{\begin{tabular}{|p{3cm}|p{3cm}|p{6cm}|p{2cm}|}
\hline
\textbf{Parameter} & \textbf{Process} & \textbf{Description} & \textbf{Default Value} \\\hline
Risks per day (RPD) & Broker & $\lambda$ value for the Poisson distribution used by the broker process to generate new risks. & 0.06  \\\hline
Number of peril regions & Broker & Number of peril regions for the catastrophes.  & 10  \\\hline
Risk limit & Broker & The maximum value of a risk. & \$10,000,000  \\\hline
Lead $top\_k$ & Broker-Syndicate Network & The number of lead syndicates a broker reaches out to. & 2 \\\hline
Follow $top\_k$ & Broker-Syndicate Network & The number of follow syndicates a broker reaches out to. & 5  \\\hline
Yearly Claim Frequency & Attritional Loss Generator & $\lambda$ value for the Poisson distribution for the number of attritional claims generated per year. & 0.1 \\\hline
$COV$ & Attritional Loss Generator & Coefficient of Variation for the gamma distribution which generates the severity of attritional claim events. & 1 \\\hline
$\mu$ & Attritional Loss Generator & Mean of the gamma distribution which generates the severity of attritional claim events. & \$3,000,000 \\\hline
Mean catastrophe events per year & Catastrophe Loss Generator & $\lambda$ value for the Poisson distribution for the number of catastrophe claims generated per year. & 0.05 \\\hline
Pareto shape & Catastrophe Loss Generator & The shape parameter of the Pareto distribution which generates the severity of catastrophe claim events. & 5 \\\hline
Minimum catastrophe damage & Catastrophe Loss Generator & The minimum value for an event to be considered a catastrophe (fraction of the risk limit). & 0.25 \\\hline
Capital & Syndicate Process & The initial syndicate capital. & \$10,000,000 \\\hline
Default lead quote line size & Syndicate Process & Default value for the lead line size offered. & 0.5 \\\hline
Default follow quote line size & Syndicate Process & Default value for the follow line size offered. & 0.1  \\\hline
Internal experience weight & Actuarial Sub-Process & Weighting which dictates whether the actuarial pricing is based primarily on syndicate history or industry-wide history. & 0.5 \\\hline
Loss experience recency weight & Actuarial Sub-Process & Weighting which dictates whether actuarial pricing weighs the past losses more than recent losses. & 0.2 \\\hline
Volatility weight ($\alpha$) & Actuarial Sub-Process & Scaling factor which dictates how much actuarial pricing considers the standard deviation of losses i.e. volatility of losses. & 0 \\\hline
Underwriter markup recency weighting & Underwriter Sub-Process & Weighting which dictates whether the underwriter markup weights past mark-ups more than recent ones. & 0.2 \\\hline
Profit fraction ($\gamma$) & Dividend Sub-Module & Fraction of profit vented as dividend. & 0.4 \\\hline
VaR EM exceedance probability ($\alpha$) & VaR EM Sub-Process & The tail probability used in the VaR calculations. & 0.05 \\\hline
VaR EM safety factor & VaR EM Sub-Process & Scaling safety factor applied to the VaR value - larger values employ more conservative EM. & 1 \\\hline
Premium reserve ratio & Premium EM Sub-Process & Premium to capital reserve ratio. & 0.5 \\\hline
Minimum capital reserving ratio & Premium EM Sub-Process & Reserved capital to working capital ratio. & 1 \\\hline
Maximum scaling factor & Premium EM Sub-Process & Minimum scaling factor applied to premium based on EM. & 1 \\\hline
\end{tabular}}
\caption{Model input parameter descriptions}
\label{tab:input_parameters}
\end{table}

\section{Results \& discussion}
\label{results_discussion-section}

The Lloyd's of London specialty insurance market is a complex system. Many in academia and the insurance industries have tried to identify the underlying processes that lead to the emergence of phenomena at the aggregate level. One particular phenomena that has been an area of focus, is the underwriting cycle (as described in Section~\ref{literature_review-section}). As a proof-of-concept we propose several experiments that aim to simulate conditions in the market with varying complexity to reproduce existing industry phenomena. Whereby practitioners and researchers can investigate the underlying features of the market that lead to these trends, better preparing those involved in the market to deal with exogenous shocks.

This section first presents empirical market dynamics, then delves into the four experiments, starting with a base case, introduction of catastrophe events, syndicates adapting to these events and finally the introduction of lead and follow dynamics (described in the previous section~\ref{model_description-section}).


\begin{figure}[!htbp]
    \centering
    \includegraphics[width=0.70\textwidth]{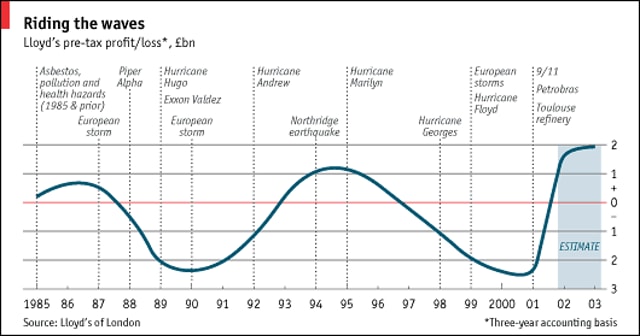}
    \caption{Aggregate Lloyd's of London market trends (see \href{https://www.economist.com/business/2013/10/03/riding-the-wave}{article})}
    \label{fig:market_dynamics_real}
\end{figure}

In Figure~\ref{fig:market_dynamics_real} the market experiences catastrophe events which exacerbates periods of hardening and softening of the market. This phenomena is conveyed as the underwriting cycle, which as a benchmark could be a phenomena we reproduce using our DES model. In the following sub-section, we describe each model scenario (experiment) and subsequently present the results while discussing the findings in relation to Lloyd's of London.

\subsection{Model scenarios}
To ensure the model is able to tell a story from simplistic behaviours to more advanced features and outcomes, we start with a base case experiment, this and subsequent experiments are described below (see Section~\ref{model_description-section} for descriptions of each model feature):

\begin{enumerate}
     \item Base case actuarial pricing, all syndicates use actuarial pricing models, premium exposure management and attritional losses occur (Scenario 1).
     \item Catastrophe event, all syndicates use actuarial pricing, premium exposure management, attritional and catastrophe losses occur (Scenario 2).
     \item Syndicates adapting to catastrophe events, all syndicates use actuarial pricing, VaR exposure management, attritional and catastrophe losses occur (Scenario 3).
     \item Leaders and Followers, syndicates can either be leaders or followers and use actuarial pricing, premium exposure management and only attritional losses occur (Scenario 4).
\end{enumerate}

\begin{table}[!htbp]
\centering
\begin{adjustbox}{width=12cm}
\begin{tabular}{|l|l|l|l|l|}
\hline
\textbf{Parameters} & \textbf{Scenario 1} & \textbf{Scenario 2} & \textbf{Scenario 3} & \textbf{Scenario 4} \\\hline
Simulation end time (years) & 50 & 50 & 50 & 50  \\\hline
RPD & 0.06 & 0.06 & 0.06 & 0.06  \\\hline
Number of peril regions & 10 & 10 & 10 & 10 \\\hline
Risk limit & 10,000,000 & 10,000,000 & 10,000,000 & 10,000,000  \\\hline
Lead $top\_k$ & 2 & 2 & 2 & 2 \\\hline
Follow $top\_k$ & - & - & - & 5 \\\hline
Yearly Claim Frequency $\lambda$ & 0.1 & 0.1 & 0.1 & 0.1 \\\hline
$COV$ & 1 & 1 & 1 & 1\\\hline
$\mu$ & 3,000,000 & 3,000,000 & 3,000,000 & 3,000,000 \\\hline
Mean catastrophe events per year & - & 0.05 & 0.05 & - \\\hline
Pareto shape & - & 5 & 5 & - \\\hline
Minimum catastrophe damage & - & 0.25 & 0.25 & - \\\hline
Capital & 10,000,000 & 10,000,000 & 10,000,000 & 10,000,000 \\\hline
Default lead quote line size & - & - & - & 0.5 \\\hline
Default follow quote line size & - & - & - & 0.1 \\\hline
Internal experience weight & 0.5 & 0.5 & 0.5 & 0.5 \\\hline
Loss experience recency weight & 0.2 & 0.2 & 0.2 & 0.2 \\\hline
Volatility weight ($\alpha$) & 0 & 0 & 0 & 0 \\\hline
VaR EM exceedance probability ($\alpha$) & - & - & 0.05 & - \\\hline
VaR EM safety factor & - & - & 1 & - \\\hline
Premium reserve ratio & 0.5 & 0.5 & - & 0.5 \\\hline
Minimum capital reserving ratio & 1 & 1 & - & 1 \\\hline
Maximum scaling factor & 1 & 1 & - & 1 \\\hline
\end{tabular}
\end{adjustbox}
\caption{Input parameter values for the four experiment scenarios}
\label{Input_Params}
\end{table}

\subsection{Base case actuarial pricing (scenario 1)}
\label{base_case_experiment_scenario_1}

This experiment utilises the most basic components of the model, in order to demonstrate the ability for core features to represent important market phenomena. In this experiment, we include five syndicates and twenty-five brokers. This maintains the 1/5 ratio of syndicates to brokers as is the case in the Lloyd's of London market pocket guide~\cite{Neal2019LloydsGuide}. As we are only utilising the actuarial pricing model, we expect the premium prices to converge around the fair price, which given the model parameters~Table~\ref{Input_Params} is \$300,000. Secondly, we expect the loss ratio of syndicates to fluctuate between periods of profitability and losses, i.e., early signs of cyclicality. Note, due to computational complexity, the model was only run for a limited number of times.

\begin{figure}[H]
     \centering
     \begin{subfigure}[b]{0.9\textwidth}
         \centering
         \includegraphics[width=\textwidth]{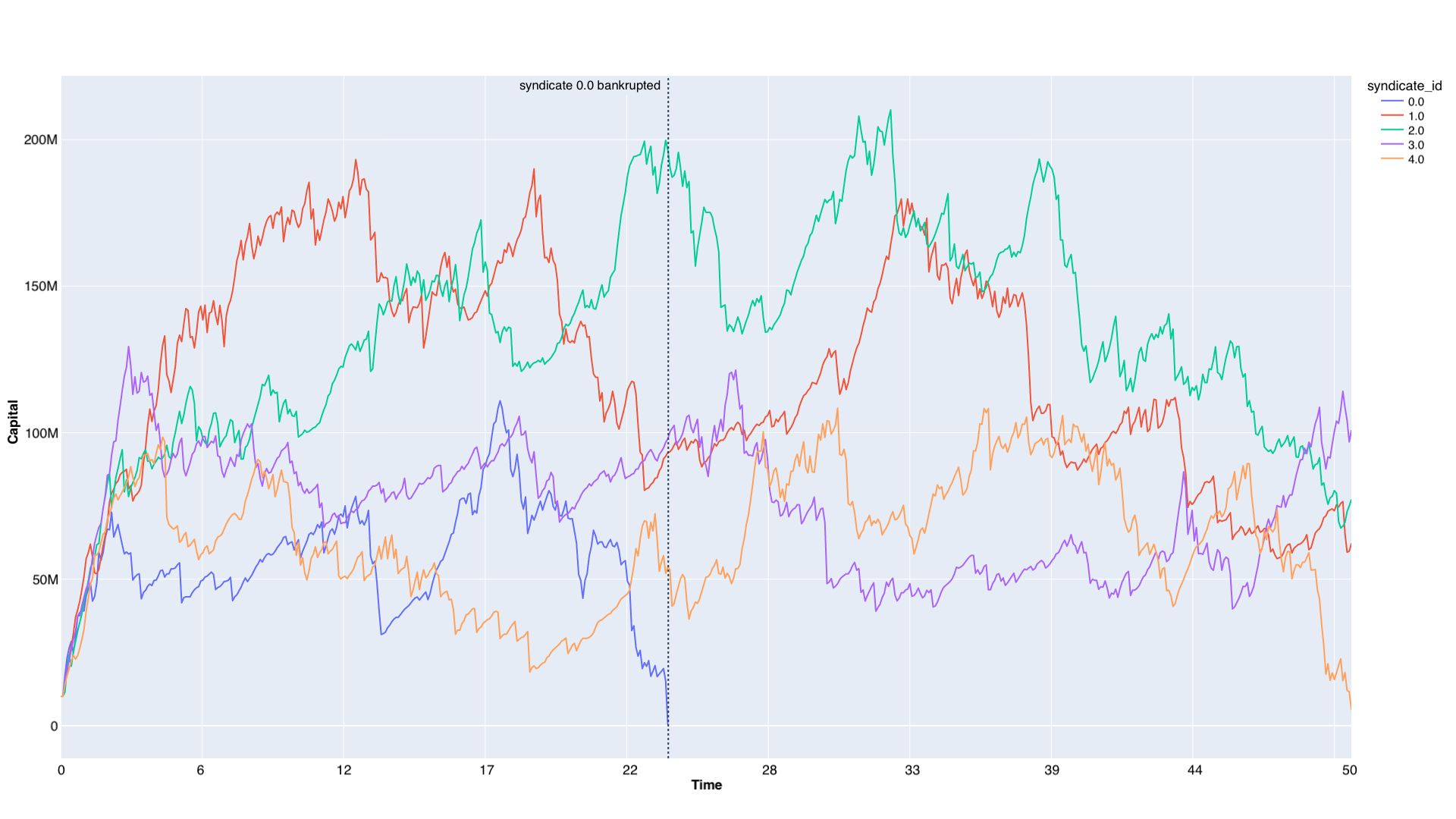}
         \caption{}
         \label{fig:capitals_scenario_1}
     \end{subfigure}
     ~
     \begin{subfigure}[b]{0.9\textwidth}
         \centering
         \includegraphics[width=\textwidth]{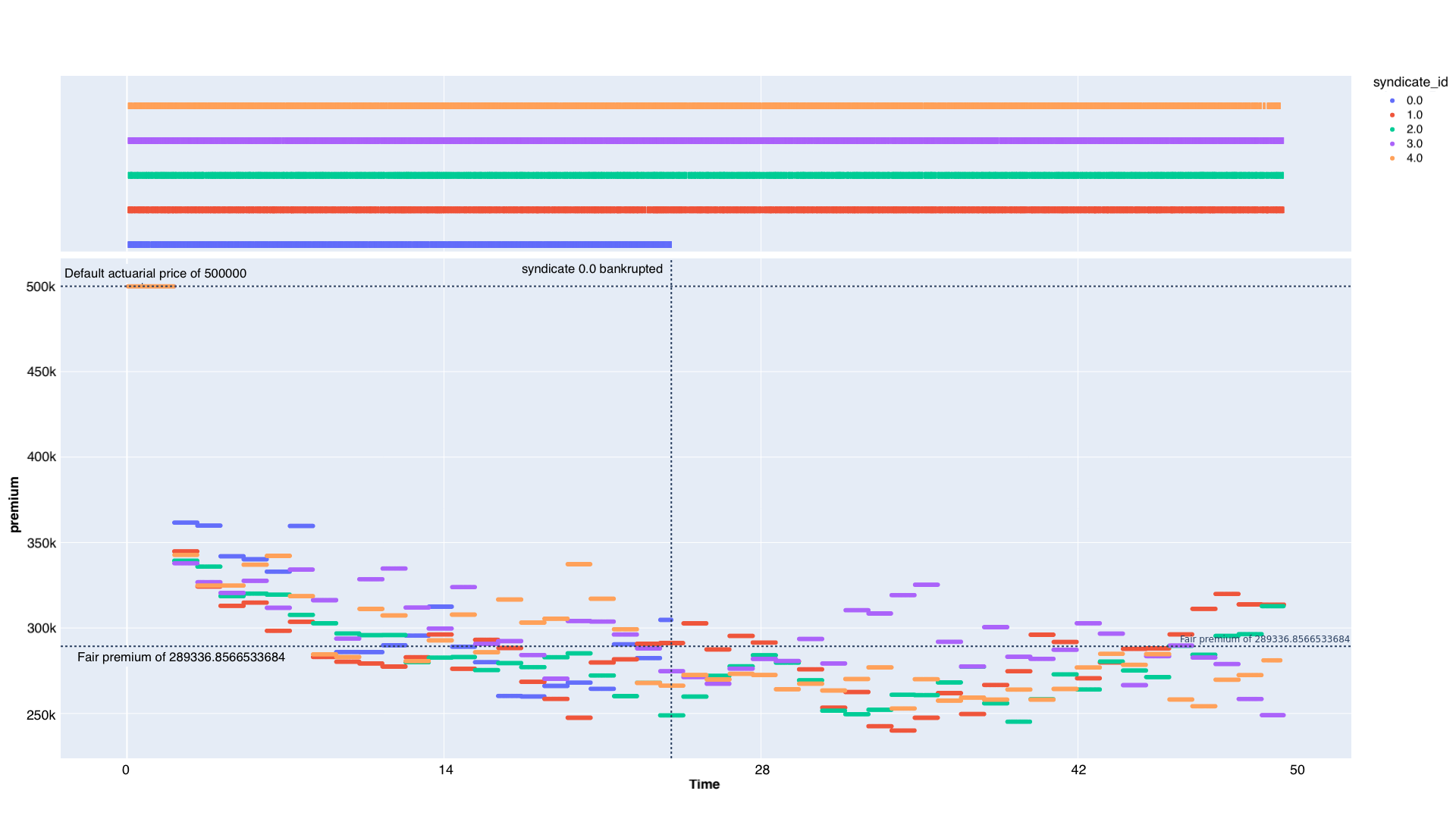}
         \caption{}
         \label{fig:premium_offered_scenario_1}
     \end{subfigure}
        \caption{Syndicate level data showing (a) capital and (b) premium offered (where time is aggregated to years).}
        \label{fig:syndicate_level_capital_premium_scenario_1}
\end{figure}

In sub-figure~\ref{fig:capitals_scenario_1}, the syndicate response over time, is indicative of capital fluctuations in reality, as observed in \cite{Heinrich2021AHomogeneity} where the authors discuss capital profiles over time. For instance, some syndicates, i.e., syndicate 0, are bankrupted as a result of their underwriting strategy, while others ride the boom and bust period better. As we hypothesised, the premiums offered have converged to the actuarial fair price of approximately \$300,000 sub-figure~\ref{fig:premium_offered_scenario_1}, the reason for this (refer to the sub-section~\ref{sub-section:actuarial}) is because the syndicate's actuarial pricing, offers a price which attempts to cover its prior losses. Given the current model setup, the mean of the prior losses is \$300,000.

\begin{figure}[!htbp]
    \centering
    \includegraphics[width=0.90\textwidth]{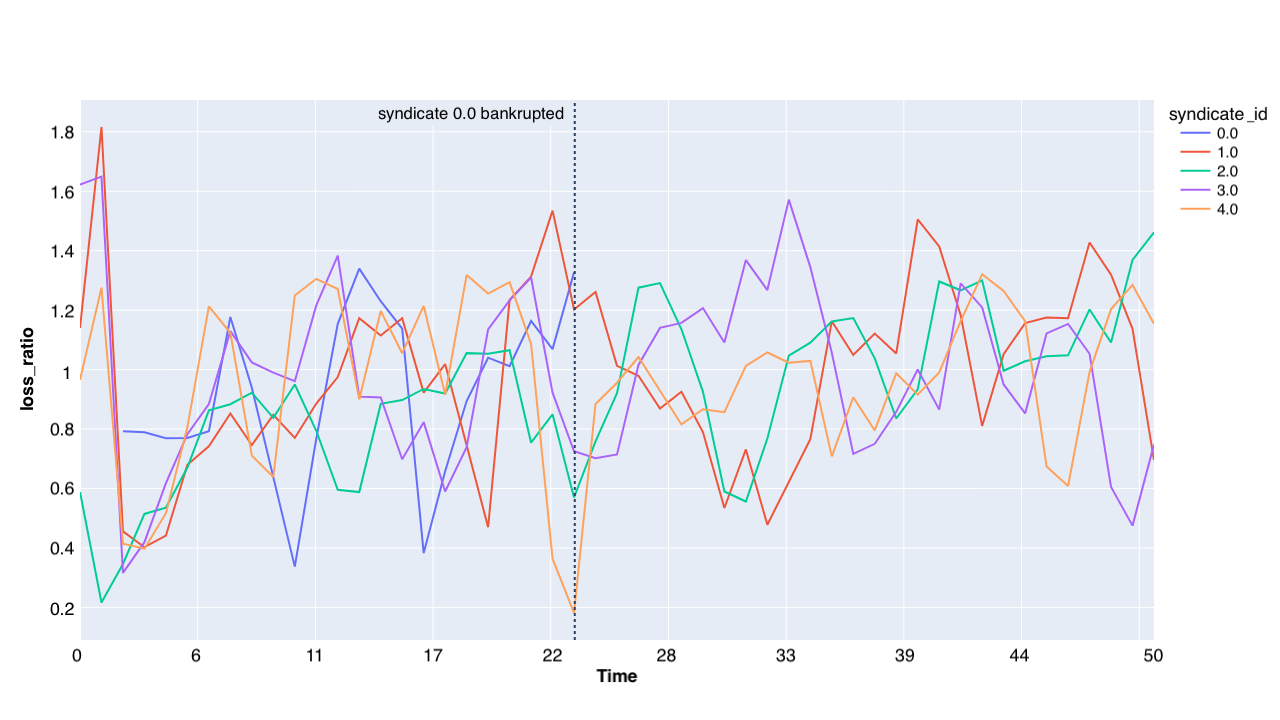}
    \caption{Syndicate level data showing loss ratios (where time is aggregated to years).}
    \label{fig:syndicate_level_loss_ratio_scenario_1}
\end{figure}

For context, the loss ratio when $\geq 1$ indicates an unprofitable syndicate as their losses are greater than their income, i.e., premiums. Conversely, $< 1$ leads to profitable syndicates. As mentioned, periods of profitability and losses is a crucial trait of all models that simulate the insurance market as described in \cite{Owadally2018TheCrises, Owadally2019AnMarkets, Zhou2013ApplicationCycles, Heinrich2021AHomogeneity}. Clearly, our model outputs also demonstrate this behaviour as observed in Figure~\ref{fig:syndicate_level_loss_ratio_scenario_1}.

\subsection{World shock events, where catastrophes meet the insurance market (scenario 2)}
\label{world_events_experiment_scenario_2}

In this experiment, we repeat the input parameters earlier, however, now we introduce the catastrophe loss generator. When catastrophe events occur, this should lead to major losses and the severity should vary across syndicates depending on their underwriting strategies. We expect to see some syndicates go insolvent, while others may not. Furthermore, insurance insiders and academics claim that the primary cause of cyclicality in the market, is due to unexpected catastrophe events \cite{boyer2012, Manikowski2012CyclicalityMarket, Venezian1985RatemakingInsurance}. Our model allows us to test this hypothesis, which we attempt in this section. These experiments should show exaggerated cyclical trends compared to the previous experiment. 

\begin{figure}[H]
     \centering
     \begin{subfigure}[b]{0.9\textwidth}
         \centering
         \includegraphics[width=\textwidth]{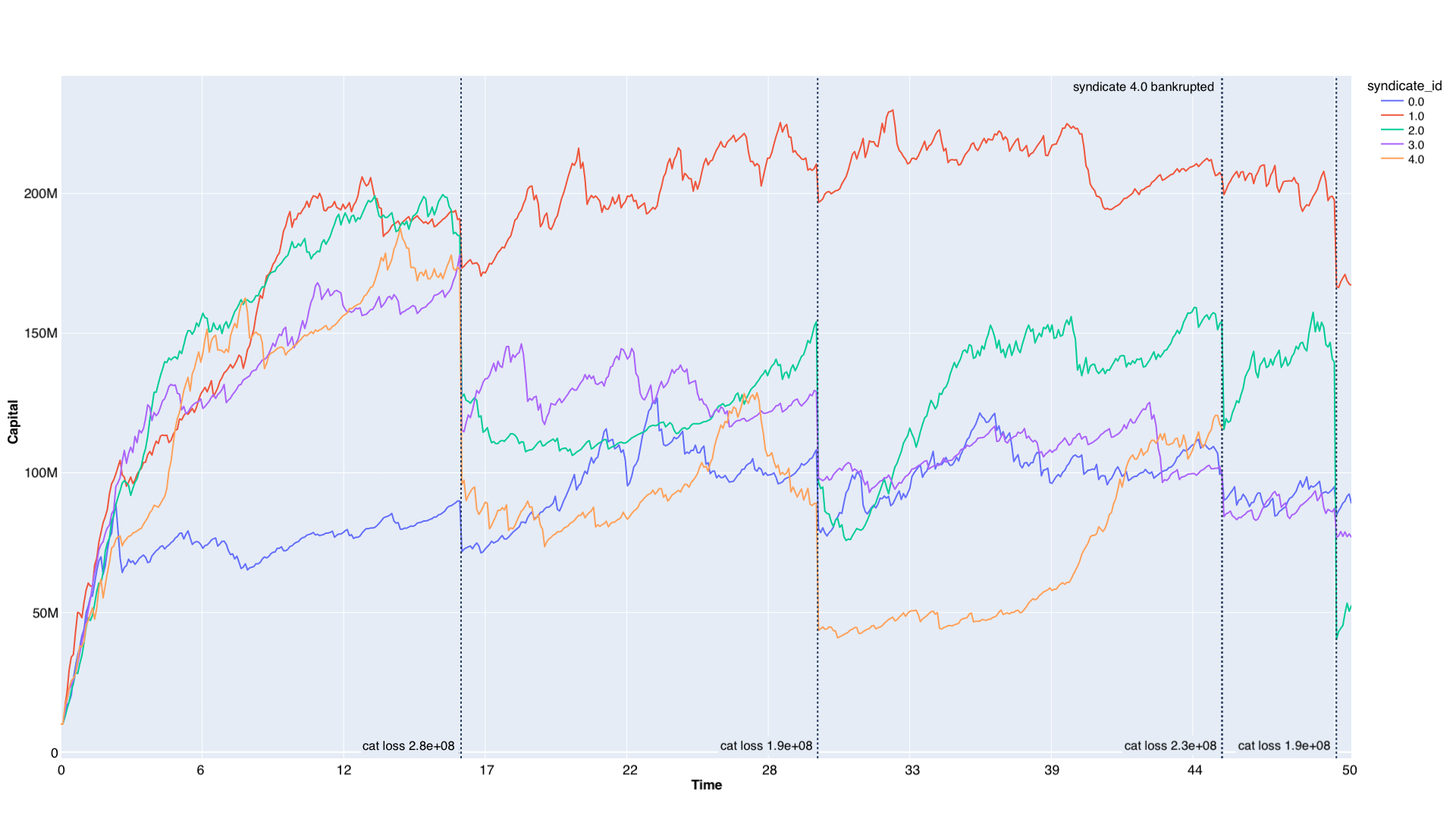}
         \caption{}
         \label{fig:capitals_scenario_2}
     \end{subfigure}
     ~
     \begin{subfigure}[b]{0.9\textwidth}
         \centering
         \includegraphics[width=\textwidth]{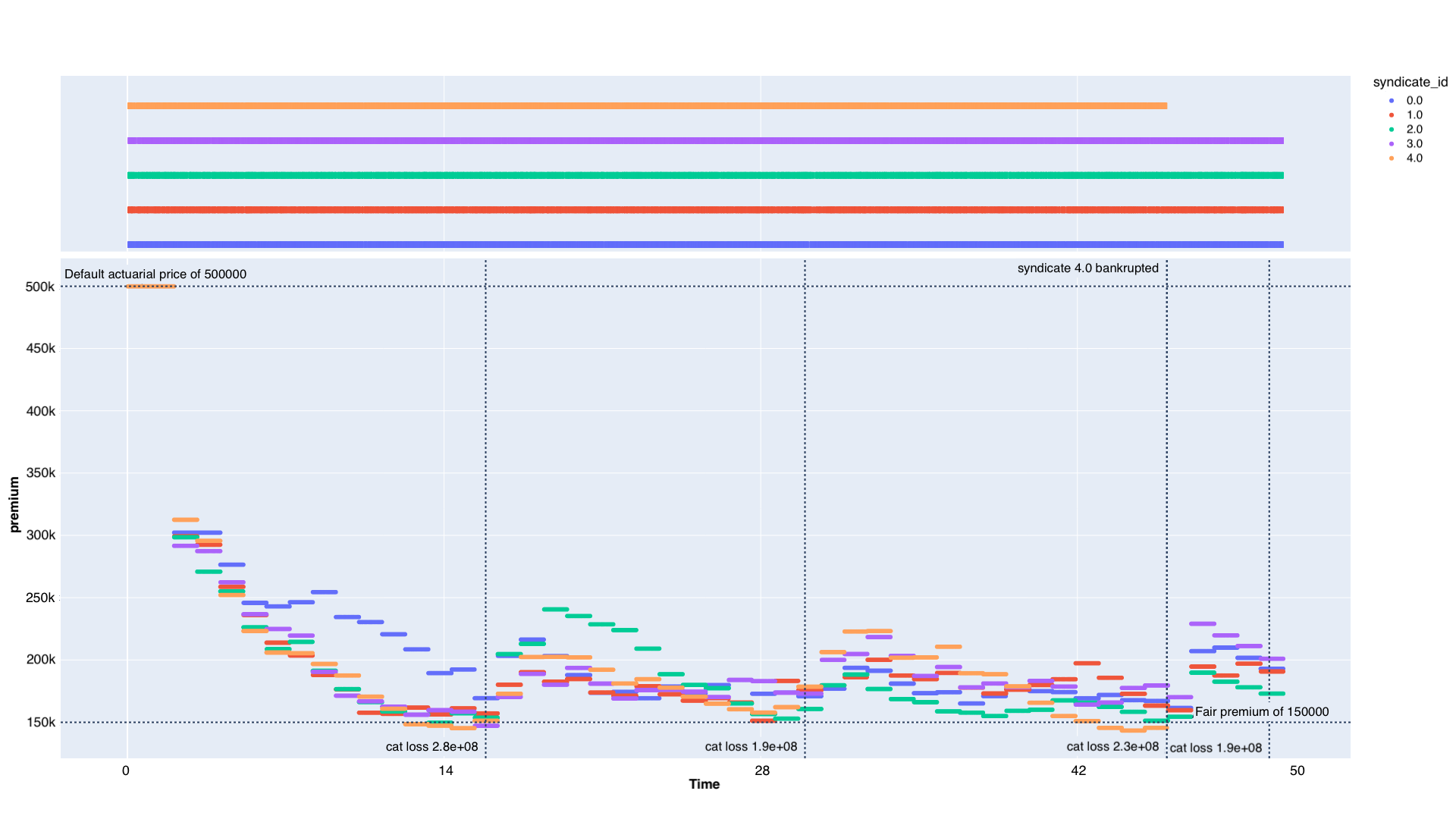}
         \caption{}
         \label{fig:premium_offered_scenario_2}
     \end{subfigure}
        \caption{Syndicate level data showing (a) capital and (b) premium offered (where time is aggregated to years).}
        \label{fig:syndicate_level_capital_premium_scenario_2}
\end{figure}

The most intriguing result from this experiment, is shown in Figure~\ref{fig:syndicate_level_capital_premium_scenario_2}, which demonstrates not only the presence of cyclicality in the premiums offered, but provides an explanation for why this phenomena occurs. Put simply, the cyclicality occurs, in the initial phase, when premiums begin to converge to a fair price, however, catastrophe events which result in large losses, forces syndicates to price premiums higher resulting in an increase in the prices offered. Eventually, once the effect of the catastrophe wears off, the syndicates once again try to converge towards the fair price and the cycle repeats itself.

\begin{figure}[H]
    \centering
    \includegraphics[width=0.90\textwidth]{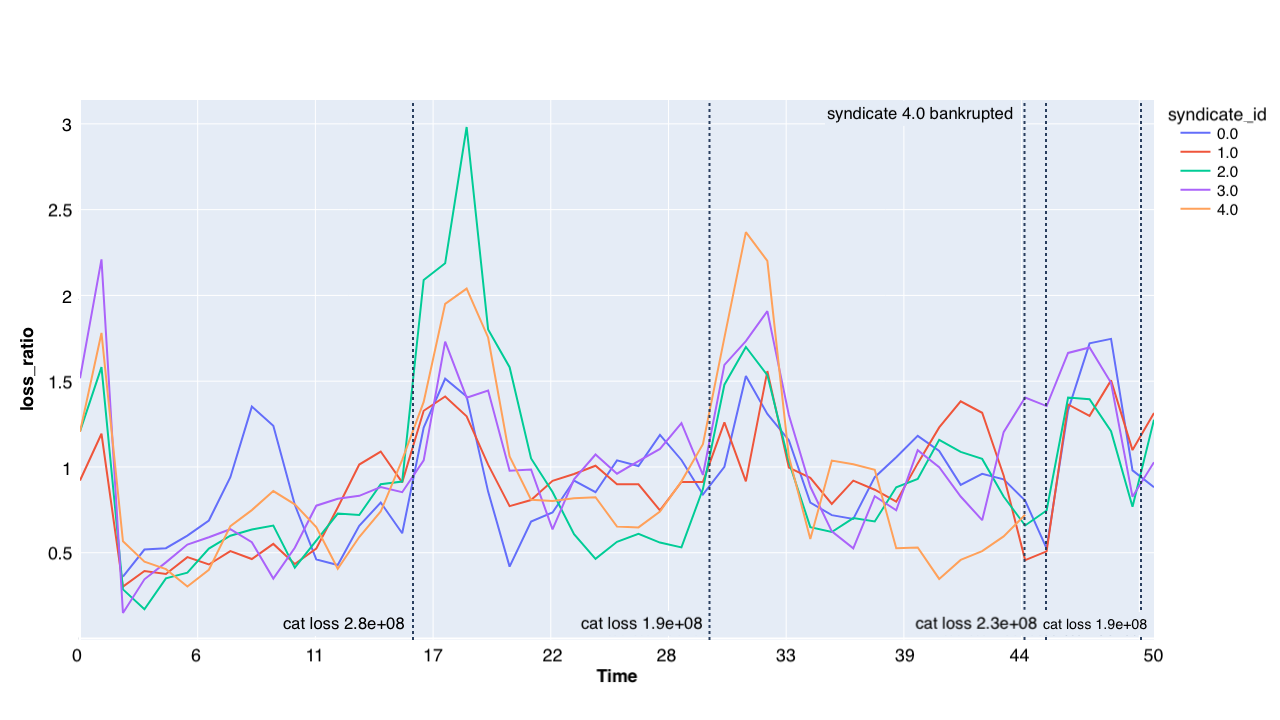}
    \caption{Syndicate level data showing loss ratios (where time is aggregated to years).}
    \label{fig:syndicate_level_loss_ratio_scenario_2}
\end{figure}

As mentioned previously, cyclicality is pronounced and in some cases, due to large catastrophe losses, loss ratios go above one Figure~\ref{fig:syndicate_level_loss_ratio_scenario_2}.

\subsection{Adapting to market shocks, the utility of VaR exposure management (scenario 3)}
\label{exposure_experiment_scenario_3}

The purpose of this experiment, is to specifically showcase the effects of advanced exposure management methods. Due to brevity, we only showcase results indicative of this objective. To this end, syndicates have utilised simplistic premium exposure management. Now, we introduce the VaR exposure management, which allows syndicates to manage their exposure in a more sophisticated manner representing behaviours of real-world syndicates more closely. 

\begin{figure}[H]
    \centering
    \includegraphics[width=0.95\textwidth]{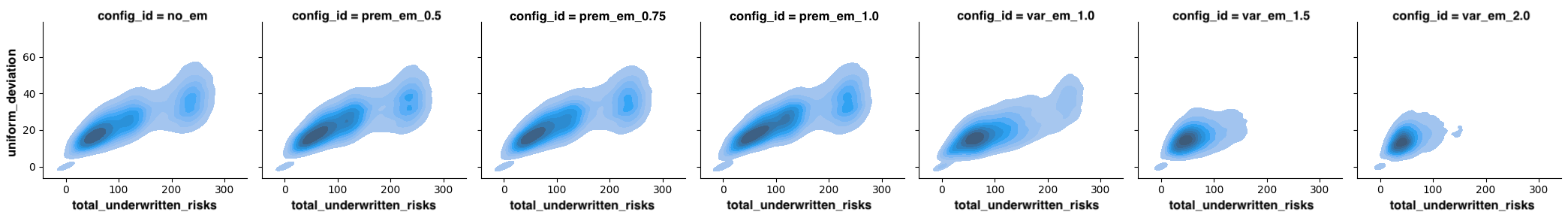}
    \caption{Uniform deviation of different exposure management processes.}
    \label{fig:uniform_deviation_exposure_scenario_3}
\end{figure}

VaR exposure management \cite{Heinrich2021AHomogeneity} should enable syndicates to adopt underwriting strategies which avoid over-subscribing to a given peril-region. A good exposure management strategy, is to distribute risk underwritten uniformly across all peril-regions \cite{Heinrich2021AHomogeneity}. For this reason, we present the uniform deviation (measures how much the real distribution of risks in the peril regions varies from a perfectly uniform peril-region distribution, where 0 is perfectly uniform). For comparison, we include no exposure management and premium exposure management outputs.  As can be seen from  Figure~\ref{fig:uniform_deviation_exposure_scenario_3}: From left to right, we observe that as the exposure management becoming more sophisticated and stringent, it is clear to see that the uniform deviation moves closer to zero.

\subsection{Leaders and followers among syndicates (scenario 4)}
\label{lead_follow_scenario_4}

The Lloyd's of London specialty insurance market as described in Section~\ref{literature_review-section} differs from other insurance markets. The most prominent difference is the syndication of risk, i.e., the ability for multiple syndicates to underwrite a risk either as leaders or followers. Lloyd's of London claims that the syndication of complex risks across multiple syndicates, allows these unconventional risks and the potential losses to be distributed among several insurers as opposed to one. This should result in less volatility, and lower likelihoods of insolvency as described in the documentation published by Lloyd's~\cite{Neal2019LloydsGuide} and~\cite{London2019BusinessMarket}. However, as far as the authors are aware, these claims have not been quantitatively tested in academic research, until now where our DES model allows us to investigate this particular claim. 

\begin{figure}[H]
     \centering
     \begin{subfigure}[b]{0.9\textwidth}
         \centering
         \includegraphics[width=\textwidth]{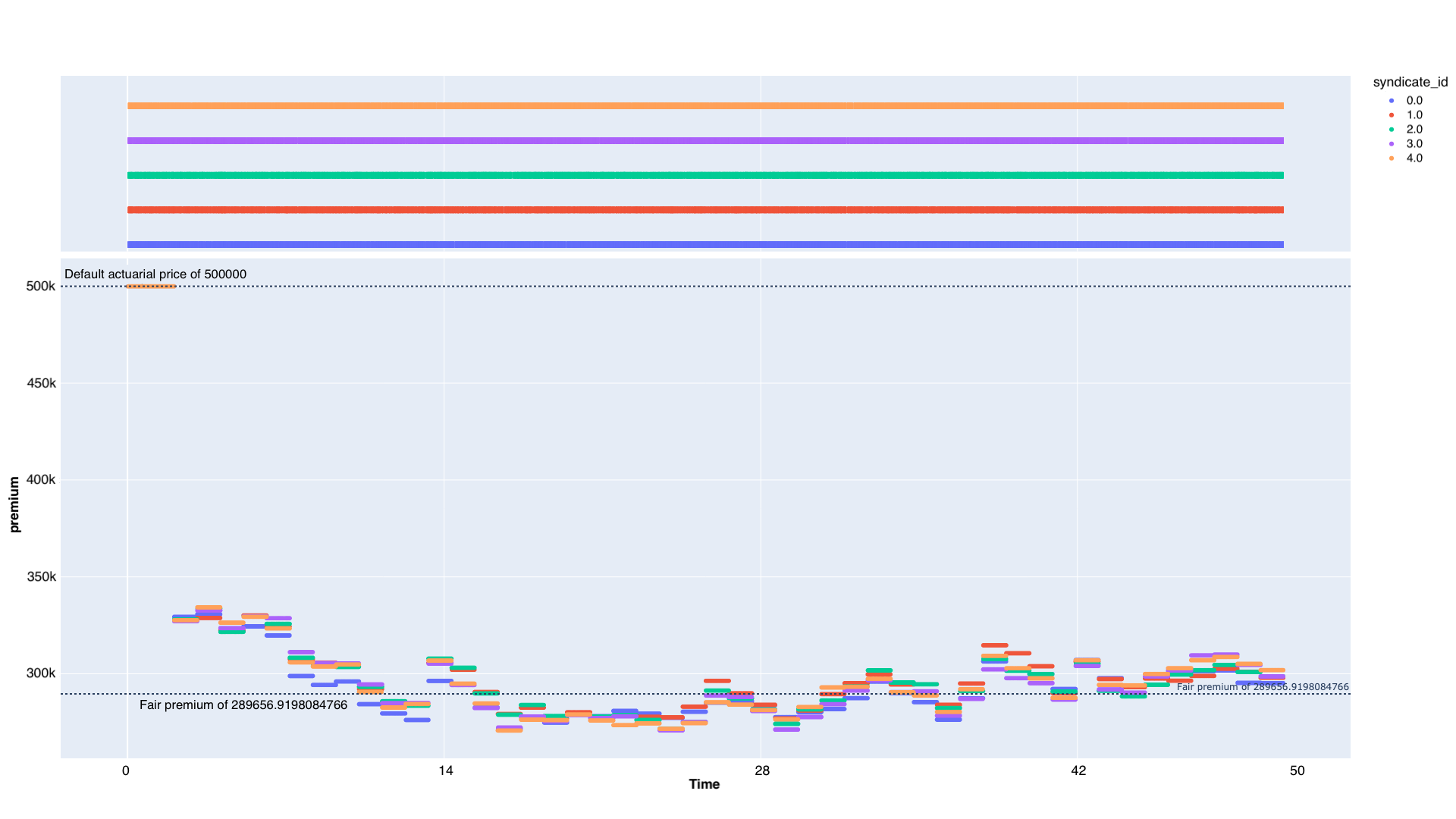}
         \caption{}
         \label{fig:premiums_scenario_4}
     \end{subfigure}
     ~
     \begin{subfigure}[b]{0.9\textwidth}
         \centering
         \includegraphics[width=\textwidth]{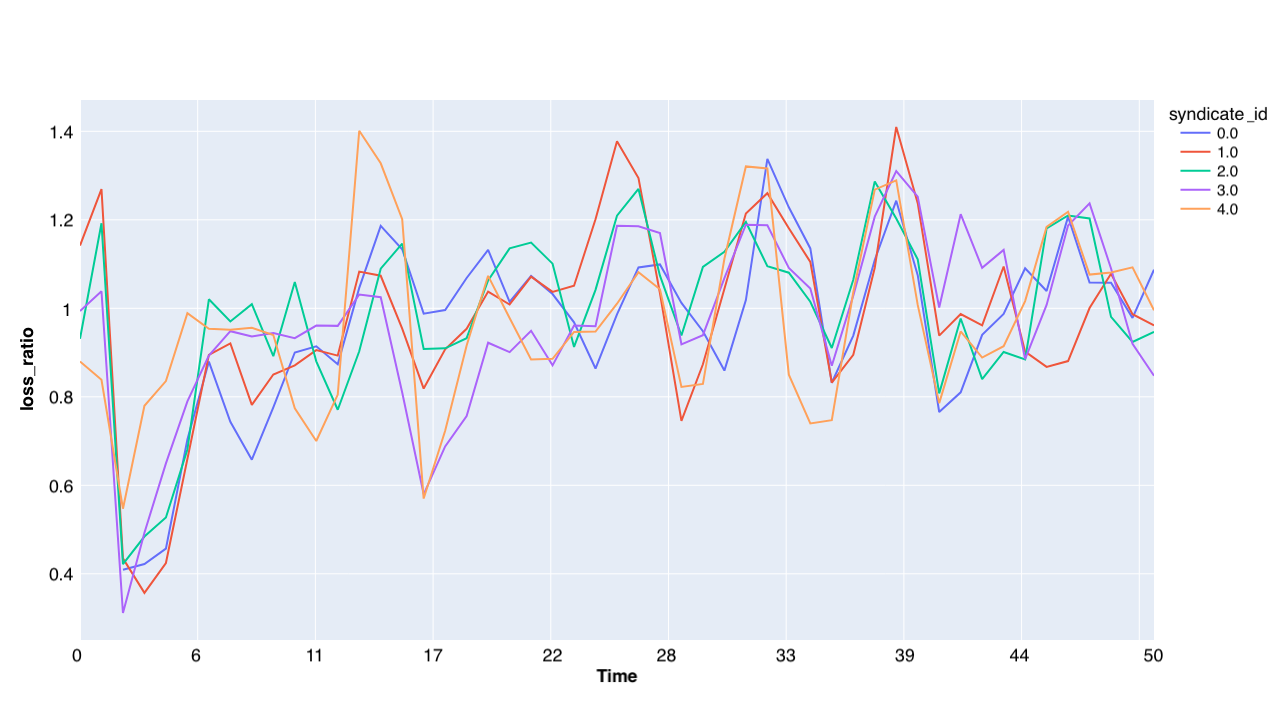}
         \caption{}
         \label{fig:loss_ratio_scenario_4}
     \end{subfigure}
        \caption{Syndicate level data showing (a) premiums offered and (b) loss ratios (where time is aggregated to years).}
        \label{fig:syndicate_level_premium_loss_ratio_scenario_4}
\end{figure}

As observed in Figure~\ref{fig:premiums_scenario_4} compared to Figure~\ref{fig:premium_offered_scenario_1} the volatility of the premium offered is significantly lower, with the premiums tightly converging towards the fair price. Due to the syndication of risks, this allows more syndicates to participate significantly in the market. This means that any losses, are shared among all syndicates, implying their loss experience is similar to each other. As a result, they all offer similar prices. This behaviour can also be observed in Figure~\ref{fig:loss_ratio_scenario_4} where the loss ratios of the syndicates are tightly coupled/correlated, indicating a similar loss experience. 

In this scenario, we observe that no insolvencies occurred, compared to the same scenario minus the lead and follow dynamics in sub-section~\ref{base_case_experiment_scenario_1}. These findings, provide strong quantitative justification for the market structure imposed by the regulator Lloyd's of London. 

In summary, the proof-of-concept DES can reproduce quantitatively and qualitatively market phenomena, such as the underwriting cycle, merits of lead - follow market structures, and the importance of exposure management.

\section{Conclusion}
\label{conclusion-section}
This research article proposes a novel DES of the Lloyd's of London specialty insurance market. The model captures granular interactions of significant actors within the marketplace, i.e., syndicates and brokers. The model has demonstrated significant results with regards to market dynamics observed in the real-world, this includes signs of cyclical behaviours congruent of the underwriting cycle, which becomes more severe with catastrophe events. Given the unique model architecture, users can swap components such as pricing models, broker-syndicate relationship networks and other features with ease that may not have been possible in past attempts such as \cite{Owadally2018TheCrises, Owadally2019AnMarkets, Zhou2013ApplicationCycles, Heinrich2021AHomogeneity}. Furthermore, the article has proposed a conceptualisation of the lead and follow syndicate dynamics which is unique to Lloyd's of London, the results from these specific experiments have reinforced the innovative market structure, employed by Lloyd's which reduces volatility in their market. 

The proposed DES has many strengths as discussed previously, for example, from a technical perspective, we adopt first-hand knowledge from experts in Lloyd's of London syndicates Ki and Brit and propose a novel, modular DES with changeable components. From a market dynamics and results perspective, we incorporate the lead and follow mechanics unique to Lloyd's of London, integrate both attritional and catastrophe loss events, and quantitatively study many important market phenomena. However, several areas of improvement can also be highlighted, for example, given the nature of the complexity which drives the market, it is difficult to abstract and quantify all aspects of the market appropriately. An example is quantifying the broker-syndicate relationship which is the cornerstone of the market. Capturing these individual-level human relationships is difficult with any modelling framework, however, individual-based modelling has allowed us to acquire new insights regarding this relationship. Furthermore, reinsurance is an important function of the insurance market. Reinsurance has a big effect on the capacity a syndicate can underwrite as they manage the tail risk of a portfolio. Modelling the availability and cost of reinsurance policies is therefore a significant driver of market dynamics. Conventional calibration and validation has not been pursued in this iteration of the proof-of-concept model, i.e., global sensitivity analysis. However, the behaviours of the model have been verified by Ki and Brit. In future there are plans to incorporate proprietary Ki and Brit data in order to calibrate the model so that these validated results can be disseminated.

Given the wealth of engineered features of the model, many future avenues can be explored. The underwriting markup (pricing model) \cite{Owadally2018TheCrises, Owadally2019AnMarkets} was not utilised in the experiments conducted in this article, however, comparing the different pricing models, i.e., actuarial and underwriting may lead to new insights with regards to market competition. Additionally, as discussed and utilised by \cite{Heinrich2021AHomogeneity}, another major component of a syndicate's activities is to pay out dividends in case of profitable performance. Given that the dividend feature is available in the model as described in Section~\ref{model_description-section} this can easily be explored as a path for future work. We hope this model provides researchers and specialty insurance practitioners with new insights that enable future R\&D projects and provide the means to reduce market volatility and enhance insurance businesses. 

\section*{Supplementary Materials}
\begin{itemize}
    \item The Hades framework is made open-source by Ki Insurance and can be found at the following link \url{https://pypi.org/project/hades-framework/}
\end{itemize}

\bibliographystyle{alpha}
\bibliography{main}

\newcommand{\etalchar}[1]{$^{#1}$}
\begin{thebibliography}{MSMM11}

\bibitem[Art06]{Arthur2006Out-of-equilibriumModeling}
W~Brian Arthur.
\newblock {Out-of-equilibrium economics and agent-based modeling}.
\newblock {\em Handbook of computational economics}, 2:1551--1564, 2006.

\bibitem[Bar99]{Barrett1999ChallengesMarket}
Michael~I Barrett.
\newblock {Challenges of EDI adoption for electronic trading in the London
  Insurance Market}.
\newblock {\em European Journal of Information Systems}, 8:1--15, 1999.

\bibitem[BFH{\etalchar{+}}16]{Baptista2016MacroprudentialMarket}
Rafa Baptista, J~Doyne Farmer, Marc Hinterschweiger, Katie Low, Daniel Tang,
  and Arzu Uluc.
\newblock {Macroprudential Policy in an Agent-Based Model of the UK Housing
  Market}.
\newblock {\em SSRN Electronic Journal}, 10 2016.

\bibitem[BJVN12]{boyer2012}
M~Martin Boyer, Eric Jacquier, and Simon Van~Norden.
\newblock {Are underwriting cycles real and forecastable?}
\newblock {\em Journal of Risk and Insurance}, 79(4):995--1015, 2012.

\bibitem[BO15]{Boyer2015UnderwritingImagination}
M~Martin Boyer and Iqbal Owadally.
\newblock {Underwriting apophenia and cryptids: Are cycles statistical figments
  of our imagination?}
\newblock {\em The Geneva Papers on Risk and Insurance-Issues and Practice},
  40:232--255, 2015.

\bibitem[CD97]{Cummins1997PriceMarkets}
J.David Cummins and Patricia~M Danzon.
\newblock {Price, Financial Quality, and Capital Flows in Insurance Markets}.
\newblock {\em Journal of Financial Intermediation}, 6(1):3--38, 1997.

\bibitem[CHT02]{Choi2002TheModels}
Seungmook Choi, Don Hardigree, and Paul~D Thistle.
\newblock {The property/liability insurance cycle: A comparison of alternative
  models}.
\newblock {\em Southern Economic Journal}, 68(3):530--548, 2002.

\bibitem[Cum92]{Cummins1992FinancialInsurance}
J~David Cummins.
\newblock {\em {Financial pricing of property and liability insurance}}.
\newblock Springer, 1992.

\bibitem[DG96]{Ding1996ModelingApproach}
Zhuanxin Ding and Clive W~J Granger.
\newblock {Modeling volatility persistence of speculative returns: a new
  approach}.
\newblock {\em Journal of econometrics}, 73(1):185--215, 1996.

\bibitem[DMF14]{DeMot2014SpecialLiability}
Jef De~Mot and Michael~G Faure.
\newblock {Special insurance systems for motor vehicle liability}.
\newblock {\em The Geneva papers on risk and insurance-issues and practice},
  39:569--584, 2014.

\bibitem[DNJH17]{Dubbelboer2017AnInsurance}
Jan Dubbelboer, Igor Nikolic, Katie Jenkins, and Jim Hall.
\newblock {An agent-based model of flood risk and insurance}.
\newblock {\em Journal of Artificial Societies and Social Simulation}, 20(1),
  2017.

\bibitem[Do00]{Dionne2000HandbookInsurance}
Georges Dionne and {others}.
\newblock {\em {Handbook of insurance}}, volume 1119.
\newblock Springer, 2000.

\bibitem[EFL10]{Einav2010BeyondMarkets}
Liran Einav, Amy Finkelstein, and Jonathan Levin.
\newblock {Beyond testing: Empirical models of insurance markets}.
\newblock {\em Annu. Rev. Econ.}, 2(1):311--336, 2010.

\bibitem[Eng02]{England2002FinancialInsurance}
Peter~D England.
\newblock {Financial Simulation Models in General Insurance}.
\newblock 2002.

\bibitem[EOW22]{England2022AnMouth}
Rei England, Iqbal Owadally, and Douglas Wright.
\newblock {An Agent-Based Model of Motor Insurance Customer Behaviour in the UK
  with Word of Mouth}.
\newblock {\em Journal of Artificial Societies and Social Simulation}, 25(2):2,
  2022.

\bibitem[FAo22]{Farmer2022Agent-BasedFuture}
J~Doyne Farmer, Robert~L Axtell, and {others}.
\newblock {Agent-Based Modeling in Economics and Finance: Past, Present, and
  Future}.
\newblock Technical report, Institute for New Economic Thinking at the Oxford
  Martin School, University{\~{}}{\ldots}, 2022.

\bibitem[Fel01]{Feldblum2001UnderwritingStrategies}
Sholom Feldblum.
\newblock {Underwriting cycles and business strategies}.
\newblock In {\em Proceedings of the Casualty Actuarial Society}, volume~88,
  pages 175--235, 2001.

\bibitem[FF09]{Farmer2009TheModelling}
J~Doyne Farmer and Duncan Foley.
\newblock {The economy needs agent-based modelling}, 2009.

\bibitem[FPC{\etalchar{+}}04]{Fitzpatrick2004BarriersThem}
Annette~L Fitzpatrick, Neil~R Powe, Lawton~S Cooper, Diane~G Ives, and John~A
  Robbins.
\newblock {Barriers to health care access among the elderly and who perceives
  them}.
\newblock {\em American journal of public health}, 94(10):1788--1794, 2004.

\bibitem[GB03]{Giardina2003BubblesModels}
Irene Giardina and J-P Bouchaud.
\newblock {Bubbles, crashes and intermittency in agent based market models}.
\newblock {\em The European Physical Journal B-Condensed Matter and Complex
  Systems}, 31:421--437, 2003.

\bibitem[Ge17]{Ge2017EndogenousMarket}
Jiaqi Ge.
\newblock {Endogenous rise and collapse of housing price: An agent-based model
  of the housing market}.
\newblock {\em Computers, Environment and Urban Systems}, 62:182--198, 10 2017.

\bibitem[GH95]{Grace1995ExternalCycle}
Martin~F Grace and Julie~L Hotchkiss.
\newblock {External impacts on the property-liability insurance cycle}.
\newblock {\em Journal of Risk and Insurance}, pages 738--754, 1995.

\bibitem[GJT19]{Groff2019StateOverview}
Elizabeth~R Groff, Shane~D Johnson, and Amy Thornton.
\newblock {State of the art in agent-based modeling of urban crime: An
  overview}.
\newblock {\em Journal of Quantitative Criminology}, 35:155--193, 2019.

\bibitem[Hal95]{Haley1995AIndustry}
Joseph~D Haley.
\newblock {A by-line cointegration analysis of underwriting margins and
  interest rates in the property-liability insurance industry}.
\newblock {\em Journal of Risk and Insurance}, pages 755--763, 1995.

\bibitem[HBY13]{Hyndman2013CoherentModels}
Rob~J Hyndman, Heather Booth, and Farah Yasmeen.
\newblock {Coherent mortality forecasting: the product-ratio method with
  functional time series models}.
\newblock {\em Demography}, 50(1):261--283, 2013.

\bibitem[HD94]{Harrington1994PriceMarkets}
Scott~E Harrington and Patricia~M Danzon.
\newblock {Price cutting in liability insurance markets}.
\newblock {\em Journal of Business}, pages 511--538, 1994.

\bibitem[HEB07]{Heppenstall2007GeneticMarket}
Alison~J Heppenstall, Andrew~J Evans, and Mark~H Birkin.
\newblock {Genetic algorithm optimisation of an agent-based model for
  simulating a retail market}.
\newblock {\em Environment and Planning B: Planning and Design}, 2007.

\bibitem[Her04]{Herschaft2004NotInsurance}
Jeremy~A Herschaft.
\newblock {Not your average coffee shop: Lloyd's of London-a
  twenty-first-century primer on the history, structure, and future of the
  backbone of marine insurance}.
\newblock {\em Tul. Mar. LJ}, 29:169, 2004.

\bibitem[HG15]{Hamill2015Agent-BasedEconomics}
Lynne Hamill and Nigel Gilbert.
\newblock {Agent-Based Modelling in Economics}.
\newblock 2015.

\bibitem[HMM01]{Hill2001ApplicationsProblems}
Raymond~R Hill, John~O Miller, and Gregory~A McIntyre.
\newblock {Applications of discrete event simulation modeling to military
  problems}.
\newblock In {\em Proceeding of the 2001 winter simulation conference (Cat. No.
  01CH37304)}, volume~1, pages 780--788, 2001.

\bibitem[HSF21]{Heinrich2021AHomogeneity}
Torsten Heinrich, Juan Sabuco, and J~Doyne Farmer.
\newblock {A simulation of the insurance industry: the problem of risk model
  homogeneity}.
\newblock {\em Journal of Economic Interaction and Coordination}, pages 1--42,
  2021.

\bibitem[HY03]{Harrington2003DoRoots}
Scott~E Harrington and Tong Yu.
\newblock {Do Property-Casualty Insurance Underwriting Margins Have Unit
  Roots?}
\newblock {\em Journal of Risk and Insurance}, 70(4):715--733, 2003.

\bibitem[IB13]{Ingram2013CollectiveTheory}
David Ingram and Elijah Bush.
\newblock {Collective approaches to risk in business: an introduction to Plural
  Rationality Theory}.
\newblock {\em North American Actuarial Journal}, 17(4):297--305, 2013.

\bibitem[IU10]{Ingram2010TheRationalities}
D~Ingram and A~Underwood.
\newblock {The human dynamics of the insurance cycle and implications for
  insurers: An introduction to the theory of plural rationalities}.
\newblock In {\em ERM Symposium/SOA Monograph}, 2010.

\bibitem[Jam07]{James2007LloydsOverview}
Julian James.
\newblock {Lloyd’s and the London insurance market: An overview}.
\newblock {\em Handbook of International Insurance: Between Global Dynamics and
  Local Contingencies}, pages 903--924, 2007.

\bibitem[JG14]{Johnson2014StrengtheningModeling}
Shane~D Johnson and Elizabeth~R Groff.
\newblock {Strengthening theoretical testing in criminology using agent-based
  modeling}.
\newblock {\em Journal of Research in Crime and Delinquency}, 51(4):509--525,
  2014.

\bibitem[JHS06]{Jacobson2006Discrete-eventSystems}
Sheldon~H Jacobson, Shane~N Hall, and James~R Swisher.
\newblock {Discrete-event simulation of health care systems}.
\newblock {\em Patient flow: Reducing delay in healthcare delivery}, pages
  211--252, 2006.

\bibitem[KN19]{Karl2019HowMarkets}
J~Bradley Karl and Charles Nyce.
\newblock {How cellphone bans affect automobile insurance markets}.
\newblock {\em Journal of Risk and Insurance}, 86(3):567--593, 2019.

\bibitem[Kun89]{Kunreuther1989TheRisks}
Howard Kunreuther.
\newblock {The role of actuaries and underwriters in insuring ambiguous risks}.
\newblock {\em Risk Analysis}, 9(3):319--328, 1989.

\bibitem[LM06]{Leng2006AnalysisInsurance}
Chao-Chun Leng and Ursina~B Meier.
\newblock {Analysis of multinational underwriting cycles in property-liability
  insurance}.
\newblock {\em The Journal of Risk Finance}, 7(2):146--159, 2006.

\bibitem[Lon19]{London2019BusinessMarket}
Lloyd's London.
\newblock {Business at Lloyd’s is still conducted face-to-face, and the
  bustling underwriting room is central to the smooth running of the market}.
\newblock Technical report, Lloyd's of London, London, 2019.

\bibitem[MSMM11]{McLane2011TheManagement}
Adam~J McLane, Christina Semeniuk, Gregory~J McDermid, and Danielle~J Marceau.
\newblock {The role of agent-based models in wildlife ecology and management}.
\newblock {\em Ecological modelling}, 222(8):1544--1556, 2011.

\bibitem[MW12]{Manikowski2012CyclicalityMarket}
Piotr Manikowski and Mary~A Weiss.
\newblock {Cyclicality or volatility? The satellite insurance market}.
\newblock {\em Space Policy}, 28(3):192--198, 2012.

\bibitem[Nea19]{Neal2019LloydsGuide}
John Neal.
\newblock {Lloyd's of London, Pocket Guide}.
\newblock Technical report, Lloyd's of London, London, 2019.

\bibitem[NMJ98]{Nevins1998AOperations}
Michael~R Nevins, Charles~M Macal, and Joseph~C Joines.
\newblock {A discrete-event simulation model for seaport operations}.
\newblock {\em Simulation}, 70(4):213--223, 1998.

\bibitem[Nor95]{Norberg1995AInsurance}
Ragnar Norberg.
\newblock {A time-continuous markov chain interest model with applications to
  insurance}.
\newblock {\em Applied stochastic models and data analysis}, 11(3):245--256,
  1995.

\bibitem[OZO{\etalchar{+}}19]{Owadally2019AnMarkets}
M~I Owadally, F~Zhou, R~Otunba, J~Lin, and I~D Wright.
\newblock {An agent-based system with temporal data mining for monitoring
  financial stability on insurance markets}.
\newblock {\em Expert Systems with Applications}, 123:270--282, 6 2019.

\bibitem[OZW18]{Owadally2018TheCrises}
Iqbal Owadally, Feng Zhou, and Douglas Wright.
\newblock {The Insurance Industry as a Complex Social System: Competition,
  Cycles and Crises}.
\newblock {\em Journal of Artificial Societies and Social Simulation}, 21(4):2,
  2018.

\bibitem[Pio16]{Piotr2016UnderwritingCrises}
Manikowski Piotr.
\newblock {Underwriting cycles and crises}.
\newblock {\em The world of the new economy}, (4):76--81, 2016.

\bibitem[Tho10]{Thoyts2010InsurancePractice}
Rob Thoyts.
\newblock {\em {Insurance theory and practice}}.
\newblock Routledge, 2010.

\bibitem[Ven85]{Venezian1985RatemakingInsurance}
Emilio~C. Venezian.
\newblock {Ratemaking Methods and Profit Cycles in Property and Liability
  Insurance}.
\newblock {\em The Journal of Risk and Insurance}, 52(3), 1985.

\bibitem[WC04]{Weiss2004U.S.Capacity}
Mary~A Weiss and Joon-Hai Chung.
\newblock {U.S. Reinsurance Prices, Financial Quality, and Global Capacity}.
\newblock {\em The Journal of Risk and Insurance}, 71(3):437--467, 2004.

\bibitem[Win94]{Winter1994TheMarkets}
Ralph~A Winter.
\newblock {The Dynamics of Competitive Insurance Markets}.
\newblock {\em Journal of Financial Intermediation}, 3(4):379--415, 1994.

\bibitem[YM20]{Yun2020HousingDebt-to-income}
Tae-Sub Yun and Il-Chul Moon.
\newblock {Housing market agent-based simulation with loan-to-value and
  debt-to-income}.
\newblock {\em Journal of Artificial Societies and Social Simulation}, 23(4),
  2020.

\bibitem[Zho13]{Zhou2013ApplicationCycles}
Feng Zhou.
\newblock {\em {Application of agent based modeling to insurance cycles}}.
\newblock PhD thesis, City University London, 2013.

\end{thebibliography}

\end{document}